\documentclass{article}
\usepackage{graphicx}
\usepackage{latexsym}
\usepackage{amsmath}
\usepackage{amssymb}
\usepackage{amsfonts}
\usepackage{amsxtra}

\allowdisplaybreaks
\newcommand{\sss}[1]{{\scriptscriptstyle #1}}

\newcommand{\Ll}{\ensuremath{\mathcal{L}}}

\newsavebox{\piracbox}
\savebox{\piracbox}{$\not\! p$}

\newsavebox{\ptiracbox}
\savebox{\ptiracbox}{$\not\! p_\sss{T}$}

\newsavebox{\qiracbox}
\savebox{\qiracbox}{$\not\! q$}

\newcommand{\Tr}[1]{\mathrm{Tr}\left( #1 \right)}

\newcommand{\mm}{\widetilde{m}}
\newcommand{\mb}{m}
\newcommand{\Mb}{M}

\newcommand{\T}{\mathcal{T}}

\newcommand{\F}{\mathcal{F}}
\renewcommand{\S}{\mathcal{S}}

\renewcommand{\P}{\mathcal{P}}

\newcommand{\xI}{I}
\newcommand{\xII}{II}
\newcommand{\xIV}{III}
\newcommand{\xjor}{NR}

\newcommand{\umax}{u_\sss{\mathrm{max}}}
\newcommand{\umin}{u_\sss{\mathrm{min}}}

\newcommand{\tmax}{t_\sss{\mathrm{max}}}
\newcommand{\tmin}{t_\sss{\mathrm{min}}}
\newcommand{\pw}[3]{\ud{f_{#1\!\!}}{#2}{#3}}
\newenvironment{sarray}{\renewcommand{\arraycolsep}{0pt}
  
  \begin{array}}{\end{array}} 
\newcommand{\uud}[4]{#1
  \begin{sarray}{ccc} {\scriptstyle #2} & {\scriptstyle #3} & \\ &  &
    {\scriptstyle #4} \end{sarray}} 
\newcommand{\udu}[4]{#1
  \begin{sarray}{ccc} {\scriptstyle #2} &  & {\scriptstyle #4}\\ &
    {\scriptstyle #3} &  \end{sarray}} 
\newcommand{\dud}[4]{#1
  \begin{sarray}{ccc} & {\scriptstyle #3}  & \\ {\scriptstyle #2} & &
    {\scriptstyle #4}  \end{sarray}} 
\newcommand{\udd}[4]{#1
  \begin{sarray}{ccc} {\scriptstyle #2} & & \\ & {\scriptstyle #3} &
    {\scriptstyle #4}  \end{sarray}} 
\newcommand{\ud}[3]{#1
  \begin{sarray}{cc}{\scriptstyle #2} &  \\ & {\scriptstyle #3}
  \end{sarray}} 
\title{Baryon polarization in low-energy unpolarized meson-baryon scattering}
\author{Antonio O.\ Bouzas \thanks{E-mail:
    abouzas@mda.cinvestav.mx}\\\small Departamento de F\'{\i}sica
  Aplicada, CINVESTAV-IPN \\\small Carretera Antigua a Progreso Km.\
  6, Apdo.\ Postal 73 ``Cordemex''\\\small
  M\'erida 97310, Yucat\'an, M\'exico}
\date{}
\begin{document}
\maketitle
\begin{abstract}
  We compute the polarization of the final-state baryon, in its rest
  frame, in low-energy meson--baryon scattering with unpolarized
  initial state, in Unitarized BChPT.  Free parameters are determined
  by fitting total and differential cross-section data (and
  spin-asymmetry or polarization data if available) for $pK^-$, $pK^+$
  and $p\pi^+$ scattering.  We also compare our results with those of
  leading-order BChPT.  
\end{abstract}

\section{Introduction}
\label{sec:intro}

The study of spin phenomena in meson--baryon low-energy scattering
provides stringent tests of QCD and its associated effective theory,
Baryon Chiral Perturbation Theory (BChPT) \cite{ber07a,sch05}.
Because mesons are spinless, and at low energies can be considered
nearly structureless, their scattering off baryons is the simplest
process from the point of view of baryon spin dynamics.  As such, it
is of great interest as a probe of, and may lead to important insights
into, the structure and dynamics of baryons.

The applicability of BChPT is limited to the near-threshold energies
at which meson momenta are much smaller than the chiral symmetry
breaking scale.  At moderately higher energies, resonances and
coupled-channels effects enter the dynamics that must be either
incorporated into the theory or dynamically generated by it.  In the
three-flavor case those phenomena may be convolved with a
strong-coupling regime originating in the large masses of strange
hadrons.  A well-known example is $N\overline{K}$ scattering, in which
several strongly-coupled channels are open at threshold, leading to a
subthreshold resonance, $\Lambda(1405)$, and rendering BChPT
inapplicable to those processes.  Many models and techniques have been
developed to overcome those difficulties over a period of several
decades, that we cannot review here.  In the specific context of
BChPT, unitary coupled-channels techniques based on
Lippmann--Schwinger or Bethe--Salpeter equations have been succesfully
applied to the study of $N\overline{K}$ and other meson--baryon
processes, even at relatively high energies
\cite{sie88,kai95,kai97,ose98,lut00,bor02}.  A unitarization method
dealing directly with the chiral effective theory $T$-matrix has been
introduced in \cite{oll99} in the meson sector, and extended to the
baryon sector in \cite{mei00,oll01,jid02}.  This Unitarized Baryon
Chiral Perturbation Theory (UBChPT) has been shown to give accurate
descriptions of unpolarized cross-section data in $N\overline{K}$
processes \cite{oll01,jid02,jid03,oll06z}, and in $N\pi$ scattering
beyond the $\Delta$ resonance peak \cite{mei00}.

In this paper we consider a particular aspect of hadron spin dynamics,
the production of polarized baryons in unpolarized meson--baryon
scattering.  Specifically, we compute the polarization of the
final-state baryon in its rest frame in low-energy two-body
meson--baryon scattering with unpolarized initial state, in UBChPT.
By low energy we mean incident-meson momentum $q_\mathrm{lab} \lesssim
300$ MeV.  We use tree-level BChPT partial waves unitarized with the
method of \cite{mei00, oll01,jid02}, and determine their free
parameters by fitting total and differential cross-section data (and
spin-asymmetry or polarization data if available) for $pK^-$, $pK^+$
and $p\pi^+$ scattering.  We also compare our results with those of a
previous leading-order BChPT calculation \cite{bou}, with the aim of
further probing its domain of applicability.

There is some unavoidable overlap with previous works (e.g.,
\cite{lut00,oll01, jid02}) since we can only fit our calculations to
the same available data as used in the previous literature.  We
remark, however, that although the unitarization method we use is the
same as in \cite{oll01,jid02,ben02}, our approximation scheme is
different because we have to include the full $s$- and $u$-channel
contributions to scattering $S$ and $P$ partial waves, as well as
baryon decuplet contributions, which are quantitatively important for
polarization observables in the energy range considered here.  By
contrast, in order to obtain a good description of unpolarized cross
sections (for, e.g., $pK^-$ scattering), it is enough to use less
detailed approximations.  Good fits to total cross sections can be
obtained by considering only $S$-wave scattering, as done at
$\mathcal{O}(q)$ in \cite{oll01} and at $\mathcal{O}(q^2)$ in
\cite{oll06z}. \footnote{ $\mathcal{O}(q^n)$ denotes a generic
  quantity of chiral order $n$, with $q$ a nominally small quantity
  such as a meson momentum or mass.} In \cite{jid02,ben02} unpolarized
differential cross sections are described in terms of $S$ and $P$
partial waves in a non-relativistic approximation in which some of the
contributions mentioned above are of subleading order, therefore
neglected.  This is to be expected, since polarized observables are
more sensitive to smaller partial waves than unpolarized ones, the
latter being usually dominated by large, resonant waves.

In the following section we present our notation and conventions, give
the explicit form of the tree-level partial waves used throughout the
paper, and very briefly discuss the unitarization method applied to
those partial waves.  In sections \ref{sec:pKm}---\ref{sec:npi} we
describe our results for $pK^-$, $pK^+$, and $p\pi^+$ scattering,
resp.  Detailed fits to scattering data are reported, and the
resulting parameters applied to the computation of final-state
polarization.  In section \ref{sec:finrem} we give some final remarks.

\section{Partial waves and unitarization}
\label{sec:pw}

The ground-state meson and baryon octets are described by standard
\cite{don94} traceless $3\times 3$ complex matrix fields $\phi$ and
$B$, resp., with $\phi$ hermitian.  We use the physical flavor basis
\begin{equation}
  \label{eq:cartanweyl}
  \begin{gathered}
    \beta^1 = \frac{1}{\sqrt{2}} \left(\lambda^1 + i \lambda^2\right),
    \quad
    \beta^2 = \beta^{1\dagger}~,
    \quad
  \beta^3 = \lambda^3~,\\
  \beta^4 = \frac{1}{\sqrt{2}} \left(\lambda^4 + i \lambda^5\right),
  \quad
  \beta^5 = \beta^{4\dagger}~,\quad  
  \beta^6 = \frac{1}{\sqrt{2}} \left(\lambda^6 + i \lambda^7\right),\quad
  \beta^7 = \beta^{6\dagger}~,\quad  
  \beta^8 = \lambda^8~,
  \end{gathered}
\end{equation}
where $\lambda^a$ are SU(3) Gell-Mann matrices. The real matrices
$\beta^a$ are not hermitian.  Their hermitian conjugates form a basis
that differs from $\{\beta^a\}_{a=1}^8$ only in its ordering.  To
distinguish field components with respect to each of those bases we
use lower flavor indices for ${\beta^\dagger}_a$.  Thus, meson and
baryon fields are decomposed as\footnote{We do not use summation
  convention for flavor indices.}  $\phi = \sum_b \phi_b \beta^b
/\sqrt{2} = \sum_b \phi^b {\beta^\dagger}_b /\sqrt{2}$ and $B = \sum_a
B_a \beta^a /\sqrt{2} = \sum_a B^a {\beta^\dagger}_a/\sqrt{2}$, with
$\phi_b = \Tr{{\beta^\dagger}_b \phi}/\sqrt{2}$, $\phi^b = \Tr{\beta^b
  \phi}/\sqrt{2}$, and similarly $B_a$ and $B^a$.  Baryon and meson
states are denoted $|B^a(p,\sigma)\rangle$ and $|M^b(q)\rangle$,
resp., with $\sigma=\pm 1/2$ the spin along a fixed spatial direction
in the fermion rest frame, and $p$, $q$ four-momenta.  We always use
hadron kets with an upper index, and bras with a lower one, with
masses $\mb_a$ and $\mm_b$ for baryons and mesons resp.  Free fields
couple to one-particle states as $\langle 0 | B_a(x) |
B^c(p,\sigma)\rangle = \delta_a^c u(p,\sigma) \exp(-ipx)$ and $\langle
0 | \phi_b(x) | M^c(q)\rangle = \delta_b^c \exp(-iqx)$.

Indices can be raised or lowered by means of the symmetric matrices
$e^{ab}=\Tr{\beta^a\beta^b}/2$ $=\Tr{{\beta^\dagger}_a
  {\beta^\dagger}_b}/2$ $= e_{ab}$ and $e^a_b = \delta^a_b$.  In this
basis the structure constants and the anticommutator constants in the
fundamental representation are
\begin{equation}
  \label{eq:fand}
  \begin{aligned}\mbox{ }
    [\beta^a,\beta^b] &= 2 \sum_c \uud{f}{a}{b}{c}\, \beta^c~,
    & & \uud{f}{a}{b}{c} = \frac{1}{4} \Tr{{\beta^\dagger}_c[\beta^a,
      \beta^b]},\\ 
    \{\beta^a,\beta^b\} &= 2 \sum_c \uud{d}{a}{b}{c}\, \beta^c +
    \frac{4}{3} e^{ab}~,
    & & \uud{d}{a}{b}{c} = \frac{1}{4} \Tr{{\beta^\dagger}_c\{\beta^a,
      \beta^b\}}.
  \end{aligned}  
\end{equation}
Similar definitions hold for $\dud{f}{a}{b}{c\,}$,
$\udd{d}{a}{b}{c}\,$, etc.  The constants $f^{abc}$ and $d^{abc}$ (as
well as $f_{abc} = - f^{abc}$ and $d_{abc} = d^{abc}$) are totally
antisymmetric and symmetric, resp.  Their numerical values are
different from their Gell-Mann--basis analogs. 

The Lagrangian of fully relativistic Baryon Chiral Perturbation Theory
(BChPT) is written as a sum $\Ll = \Ll_M + \Ll_{MB}$ of a purely
mesonic Lagrangian $\Ll_M$ and a meson--baryon one $\Ll_{MB}$.  The
mesonic Lagrangian to $\mathcal{O}(q^4)$ was first obtained
in \cite{gas84,gas85a}.  The meson--baryon Lagrangian $\Ll_{MB}$ has
been given to $\mathcal{O}(q^3)$ in the three-flavor case in
\cite{kra90,fri04,oll06a}, and in \cite{gas88} for two flavors.  The
tree-level amplitudes from $\Ll$ for meson-baryon scattering have been
given in \cite{oll01} (see also \cite{jid02,bou}).  We discuss here
the associated partial waves, as used below to fit data.  Some, but
not all, of the  $S$ and $P$ partial waves have been given explicitly
before \cite{jid02}.

Defining the $T$-matrix as $S=I + i (2\pi)^4 \delta(P_f-P_i) T$, the
scattering amplitudes are given by $T$-matrix elements
$\ud{\T}{ab}{a'b'}(s,u;\sigma,\sigma') \equiv \langle
B_{a'}(p',\sigma') M_{b'}(q') | T | B^{a}(p,\sigma) M^b(q) \rangle$ as
functions of the Mandelstam invariants $s=(p+q)^2$, $u = (p-q')^2$ and
the spin variables.  The center-of-mass (CM) frame partial waves
$\pw{\ell\pm}{ab}{a'b'}$ corresponding to $j=\ell\pm 1/2$ are defined
as,
\begin{equation}
  \label{eq:pw}
  \begin{split}
  \ud{\T}{ab}{a'b'}(s,u;\sigma,\sigma') &= \sum_{\ell = 0}^\infty
  \left\{ 
  \left((\ell+1) \pw{\ell+}{ab}{a'b'} + \ell \pw{\ell-}{ab}{a'b'}
  \right) 
  P_\ell(\hat{p}\cdot \hat{p}') \chi'^\dagger_{\sigma'} \cdot
  \chi_{\sigma} \right.\\
  &+ \left. i \left(\pw{\ell+}{ab}{a'b'} - \pw{\ell-}{ab}{a'b'}
  \right) P'_\ell(\hat{p}\cdot \hat{p}') \chi'^\dagger_{\sigma'} \cdot
  (\vec{\sigma} \cdot \hat{p}\wedge \hat{p}')\chi_{\sigma}  \right\}~,
  \end{split}
\end{equation}
with $\hat{p}\cdot \hat{p}' = \cos\theta_{CM}$, $P_\ell$ and $P'_\ell$
the Legendre polynomial of order $\ell$ and its derivative, and
$\chi_\sigma$, $\chi'_{\sigma'}$ 2-component spinors for the initial
and final baryon, resp.  With this definition the partial waves for
the contact interaction amplitude are, explicitly writing flavor
coefficients, 
\begin{equation}
  \label{eq:pwc}
  \begin{gathered}
  \ud{\F_{(c)}}{ab}{a'b'} = \sum_d \dud{f}{b'}{b}{d} \uud{f}{d}{a}{a'}~,  \\
  \pw{(c)_0}{ab}{a'b'} = \frac{\ud{\F_{(c)}}{ab}{a'b'}}{f^2} N_{a'}
  N_a \left( \sqrt{s} - \frac{\mb_a + \mb_{a'}}{2}\right),
  \quad
  \pw{(c)_{1-}}{ab}{a'b'} = \frac{\ud{\F_{(c)}}{ab}{a'b'}}{f^2}
  \frac{|\vec{p}\,'| |\vec{p}\,|}{N_{a'}N_a} \left( \sqrt{s} +
    \frac{\mb_a + \mb_{a'}}{2}\right),
  \end{gathered}
\end{equation}
with $N_{a} = \sqrt{p^0 + \mb_a}$, and $f$ the common pseudoscalar
octet decay constant.  For the $s$-channel amplitude we have,
\begin{equation}
  \label{eq:pws}
  \begin{gathered}
    \udu{\F_{(s)}}{ab}{a'b'}{\{d\}} = \left( D\uud{d}{b}{a}{d} +
      F\uud{f}{b}{a}{d} \rule{0pt}{10pt}\right)
    \left( D\udd{d}{d}{b'}{a'} - F\udd{f}{d}{b'}{a'}\right) ,\\
    \pw{(s)_0}{ab}{a'b'} = -\frac{1}{f^2} N_{a'} N_a
    (\sqrt{s}-\mb_{a'}) (\sqrt{s}-\mb_{a}) \sum_{d=1}^8
    \frac{\udu{\F_{(s)}}{ab}{a'b'}{\{d\}}}{\sqrt{s}+\mb_d}~,\\
    \pw{(s)_{1-}}{ab}{a'b'} = -\frac{1}{f^2} \frac{|\vec{p}\,'|
      |\vec{p}\,|}{N_{a'} N_a} (\sqrt{s}+\mb_{a'}) (\sqrt{s}+\mb_{a})
    \sum_{d=1}^8 \frac{\udu{\F_{(s)}}{ab}{a'b'}{\{d\}}}{\sqrt{s}-\Mb_d}~,
\end{gathered}
\end{equation}
where $M_a$ is the bare mass of the exchanged baryon of flavor $a$,
and $D$, $F$ are the leading-order baryon--meson coupling constants in
SU(3) BChPT \cite{kra90,fri04,oll06a}.  $f_{(c){\ell\pm}}$ and
$f_{(s){\ell\pm}}$ vanish for all other values of $\ell$, $j$.  For
the $u$-channel partial waves we need to introduce some notations
besides the corresponding flavor factor,
\begin{equation}
  \label{eq:notu}
  \begin{gathered}
      \udu{\F_{(u)}}{ab}{a'b'}{\{d\}} =
    \left( D\udd{d}{a}{b'}{d}   - F\udd{f}{a}{b'}{d}\right)
    \left( D\uud{d}{b}{d}{a'} + F\uud{f}{b}{d}{a'}\right)~,\quad
    \gamma_{0d} = \mb_{a'}+\mb_a-\mb_d~, \\
    \gamma_{1d} = (\mb_{a'}+\mb_{d})(\mb_{a}+\mb_{d})~,\quad
    z_d = \frac{2 \mb_d^2 - (\umax+\umin)}{\umax - \umin}~,
  \end{gathered}
\end{equation}
where the kinematic limits for $u$, $\umin \leq u \leq \umax$, are
given as functions of $s$ in appendix \ref{sec:kin}.  The $S$ wave is
then writtens as,
\begin{equation}
  \label{eq:Su}
  \begin{split}
    \pw{(u)_0}{ab}{a'b'} &= \frac{N_{a'} N_a}{f^2} \sum_{d=1}^8
    \udu{\F_{(u)}}{ab}{a'b'}{\{d\}} \left\{\sqrt{s} + \mb_d -
      \gamma_{1d} \frac{\sqrt{s}+\gamma_{0d}}{2 N_{a'}^2 N_a^2}
    \right.\\
    &\left. + \gamma_{1d} \left( -2 \frac{\sqrt{s}-\gamma_{0d}}{\umax
          - \umin} + \frac{\sqrt{s}+\gamma_{0d}}{2 N_{a'}^2 N_a^2} z_d
      \right) Q_0(z_d)\right\}~.
  \end{split}
\end{equation}
In this equation, and in what follows, $Q_\ell (z)$ denotes the
Legendre function of the second kind \cite{abram}, analytic on the $z$
plane cut along $-1< z < 1$ for $\ell$ a nonnegative integer.  In this
respect, we notice that $z_d > 1$ always, since octet baryons are
stable under strong interactions.  For the $P$-waves we have,
\begin{equation}
  \label{eq:Pu}
  \begin{aligned}
  \pw{(u)_{1-}}{ab}{a'b'} &= \frac{1}{f^2}\sum_{d=1}^8
   \udu{\F_{(u)}}{ab}{a'b'}{\{d\}}  \left\{
     \frac{|\vec{p}\,'||\vec{p}\,|}{N_{a'}N_{a}}(\sqrt{s}-m_d) +
     \frac{\gamma_{1d}(\sqrt{s}+\gamma_{0d})}{3N_{a'}N_a}
     \left( Q_2(z_d) - Q_0(z_d)\rule{0pt}{12pt}\right)
     \rule{0pt}{16pt}\right.\\
     &\left. -\frac{N_{a'}N_a}{2|\vec{p}\,'||\vec{p}\,|} \gamma_{1d}
   Q_1(z_d) \left( -\sqrt{s} + \gamma_{0d} +
     \frac{\sqrt{s}+\gamma_{0d}}{2 N_{a'}^2 N_{a}^2} 
     \left(m_d^2 - \frac{\umax + \umin}{2}\right)
     \right) \rule{0pt}{16pt}\right\},\\
  \pw{(u)_{1+}}{ab}{a'b'} &= \frac{1}{f^2}\sum_{d=1}^8
  \udu{\F_{(u)}}{ab}{a'b'}{\{d\}}  \left\{
    \frac{N_{a'}N_a}{2|\vec{p}\,'||\vec{p}\,|}(\sqrt{s}-\gamma_{0d})
    \gamma_{1d} Q_1(z_d)
   \rule{0pt}{16pt}\right.\\
    &\left. - \frac{\gamma_{1d}Q_1(z_d)}{4|\vec{p}\,'||\vec{p}\,|}
    \frac{\sqrt{s}+\gamma_{0d}}{N_{a'} N_a} \left( m_d^2 - \frac{\umax
        + \umin}{2} \right)
    - \frac{\gamma_{1d}(\sqrt{s}+\gamma_{0d})}{6N_{a'}N_a} \left(
      Q_2(z_d) - Q_0(z_d)\rule{0pt}{12pt}\right)
    \rule{0pt}{16pt}\right\}.
  \end{aligned}
\end{equation}
Finally, for the higher $u$-channel partial waves, $\ell > 1$, we have,
\begin{equation}
  \label{eq:highu}
  \begin{aligned}
   \pw{(u)_{\ell-}}{ab}{a'b'} &= \frac{1}{f^2}\sum_{d=1}^8
   \udu{\F_{(u)}}{ab}{a'b'}{\{d\}} (-1)^\ell \left\{
     \frac{\gamma_{1d}Q_\ell(z_d)}{2 |\vec{p}\,'| |\vec{p}\,|}
     \left(-N_{a'} N_{a} (\sqrt{s} - \gamma_{0d}) +
       \frac{\sqrt{s}+\gamma_{0d}}{2N_{a'} N_{a}} \times \right.
   \right. \\
   &\times \left.\left(
       m_d^2 - \frac{\umax+\umin}{2} \right)\right)
     - \left. \frac{\ell+1}{2\ell+1}
     \frac{\gamma_{1d}(\sqrt{s}+\gamma_{0d})}{2N_{a'} N_{a}} \left(
       Q_{\ell+1}(z_d) - Q_{\ell - 1}(z_d)\rule{0pt}{12pt}\right)
   \right\},\\
   \pw{(u)_{\ell+}}{ab}{a'b'} &= \frac{1}{f^2}\sum_{d=1}^8
   \udu{\F_{(u)}}{ab}{a'b'}{\{d\}} (-1)^\ell \left\{
     \frac{\gamma_{1d}Q_\ell(z_d)}{2 |\vec{p}\,'| |\vec{p}\,|}
     \left(-N_{a'} N_{a} (\sqrt{s} - \gamma_{0d}) +
       \frac{\sqrt{s}+\gamma_{0d}}{2N_{a'} N_{a}} \times \right.
   \right. \\
   &\times \left.\left(
       m_d^2 - \frac{\umax+\umin}{2} \right)\right)
     + \left. \frac{\ell}{2\ell+1}
     \frac{\gamma_{1d}(\sqrt{s}+\gamma_{0d})}{2N_{a'} N_{a}} \left(
       Q_{\ell+1}(z_d) - Q_{\ell - 1}(z_d)\rule{0pt}{12pt}\right)
   \right\}.
  \end{aligned}
\end{equation}

We turn next to decuplet baryon interactions. Those have been the
subject of an enormous literature that we cannot review here.  Our
starting point is the relativistic Lagrangian for $\Delta$-$N$-$\pi$
interaction from \cite{ber95} (see also, e.g.,
\cite{ben89,ols78,pas07}). At leading chiral order the transition from
two to three flavors amounts to inserting in the amplitudes the flavor
factors for the coupling of two octets and a decuplet.  The tree-level
scattering amplitudes so obtained differ from those for $N\pi$
scattering \cite{mei00} only in their flavor coefficients.  The CM
frame partial waves for tree-level $s$-channel decuplet exchange are,
\begin{align} 
   \pw{(s,\mathrm{dec})_{0}}{ab}{a'b'} &=  -\frac{3}{32}
  \frac{g_{10}^2 (D+F)^2}{f^2} \sum_{C=1}^{10}
   \frac{\S_{a'b'C} \ud{\S}{ab}{C}}{s^{3/2}\mb_C^2}
   \sqrt{(\sqrt{s}+\mb_{a'})^2-\mm_{b'}^2}
   \sqrt{(\sqrt{s}+\mb_{a})^2-\mm_{b}^2}\times \nonumber\\
   &\times \left\{ \sqrt{s}
     \left( (2\kappa-1)s - 2 \kappa \sqrt{s} \mb_{a'} + \mb_{a'}^2 - \mm_{b'}^2
       \rule{0pt}{12pt}\right)
     \left( (2\kappa-1)s - 2 \kappa \sqrt{s} \mb_a + \mb_a^2 - \mm_b^2
       \rule{0pt}{12pt}\right)  +
     \rule{0pt}{16pt}\right. \nonumber\\
   &+ \mb_C \left[ 4 \kappa (1-2\kappa) s^2 + (8\kappa^2 - 4\kappa
     +1) (\mb_{a'} + \mb_a) s^{3/2} \rule{0pt}{14pt}\right. \nonumber\\
   &+ (\mm_{b'}^2 + \mm_b^2 -\mb_{a'}^2
   -\mb_a^2 + 4\kappa (1-2\kappa) \mb_{a'} \mb_a) s \nonumber\\
   &\left.\left. + (-\mb_{a'} \mb_a^2 - \mb_{a'}^2 \mb_a + \mb_a \mm_{b'}^2 +
   \mb_{a'} \mm_{b}^2) \sqrt{s}  + 2 (\mb_{a'}^2 - \mm_{b'}^2)
   (\mb_a^2 - \mm_b^2) \rule{0pt}{14pt} \right]
   \rule{0pt}{16pt}\right\}, \nonumber\\     
   \pw{(s,\mathrm{dec})_{1-}}{ab}{a'b'} &=  -\frac{3}{32}
  \frac{g_{10}^2 (D+F)^2}{f^2} \sum_{C=1}^{10}
   \frac{\S_{a'b'C} \ud{\S}{ab}{C}}{s^{3/2}\mb_C^2}
   \sqrt{(\sqrt{s}-\mb_{a'})^2-\mm_{b'}^2}
   \sqrt{(\sqrt{s}-\mb_{a})^2-\mm_{b}^2}\times \nonumber\\
   &\times \left\{ \sqrt{s}
     \left( (2\kappa-1)s + 2 \kappa \sqrt{s} \mb_{a'} + \mb_{a'}^2 - \mm_{b'}^2
       \rule{0pt}{12pt}\right)
     \left( (2\kappa-1)s + 2 \kappa \sqrt{s} \mb_a + \mb_a^2 - \mm_b^2
       \rule{0pt}{12pt}\right)  +
     \rule{0pt}{16pt}\right. \nonumber\\
   & + \mb_C \left[ -4 \kappa (1-2\kappa) s^2 + (8\kappa^2 - 4\kappa
     +1) (\mb_{a'} + \mb_a) s^{3/2} \rule{0pt}{14pt}\right. \nonumber\\
   &- (\mm_{b'}^2 + \mm_b^2 -\mb_{a'}^2
   -\mb_a^2 + 4\kappa (1-2\kappa) \mb_{a'} \mb_a) s \nonumber\\
   &\left.\left. + (-\mb_{a'} \mb_a^2 - \mb_{a'}^2 \mb_a + \mb_a \mm_{b'}^2 +
   \mb_{a'} \mm_{b}^2) \sqrt{s}  - 2 (\mb_{a'}^2 - \mm_{b'}^2)
   (\mb_a^2 - \mm_b^2) \rule{0pt}{14pt} \right]
   \rule{0pt}{16pt}\right\},  \label{eq:decu1p}\\     
   \pw{(s,\mathrm{dec})_{1+}}{ab}{a'b'} &=  -\frac{3}{64}
   \frac{g_{10}^2 (D+F)^2}{f^2} \sum_{C=1}^{10} \frac{\S_{a'b'C}
     \ud{\S}{ab}{C}}{s^{3/2} (\sqrt{s}-\Mb_C)}
   \left((\sqrt{s}+\mb_{a'})^2-\mm_{b'}^2 \rule{0pt}{12pt}\right)
   \left((\sqrt{s}+\mb_{a})^2-\mm_{b}^2 \rule{0pt}{12pt}\right)\times
   \nonumber\\
   &\times  \sqrt{(\sqrt{s}-\mb_{a'})^2-\mm_{b'}^2}
   \sqrt{(\sqrt{s}-\mb_{a})^2-\mm_{b}^2}~, \nonumber\\
   \pw{(s,\mathrm{dec})_{2-}}{ab}{a'b'} &= -\frac{3}{64}
   \frac{g_{10}^2 (D+F)^2}{f^2} \sum_{C=1}^{10} \frac{\S_{a'b'C}
     \ud{\S}{ab}{C}}{s^{3/2} (\sqrt{s}+\Mb_C)}
   \left((\sqrt{s}-\mb_{a'})^2-\mm_{b'}^2 \rule{0pt}{12pt}\right)
   \left((\sqrt{s}-\mb_{a})^2-\mm_{b}^2 \rule{0pt}{12pt}\right)\times
   \nonumber\\
   &\times \sqrt{(\sqrt{s}+\mb_{a'})^2-\mm_{b'}^2}
   \sqrt{(\sqrt{s}+\mb_{a})^2-\mm_{b}^2}~.  \nonumber
\end{align}
All other partial waves $\pw{(s,\mathrm{dec})_{\ell\pm}}{ab}{a'b'}$
with $\ell\geq 2$ vanish.  In (\ref{eq:decu1p}) capital Roman letters
are used for flavor decuplet indices. $M_C$ denotes the bare mass of
the $C^\mathrm{th}$ decuplet state.  The flavor coefficients for the
initial and final vertices in $s$-channel diagrams are,
\begin{equation}
  \label{eq:flavor10}
  \ud{\S}{ab}{C} = \frac{1}{2} \varepsilon_{ilm} (\beta^a)_{jl}
  (\beta^b)_{km} (T_C)_{ijk}~,
  \quad
  \S_{a'b'C} = \frac{1}{2} \varepsilon_{ilm} (\beta^\dagger_{a'})_{lj}
  (\beta^\dagger_{b'})_{mk} (T_C)_{ijk}~,
\end{equation}
with repeated indices $i$, $j$, \ldots, summed from 1 to 3, and the
matrices $\beta$ from (\ref{eq:cartanweyl}).  $T_C$ in
(\ref{eq:flavor10}) are a standard basis for the decuplet
representation space (as given, e.g., in eq.\ (9) of \cite{leb94}).  A
coupling constant $g_{10}$ has been introduced in the partial
waves (\ref{eq:decu1p}), allowing for departures from the
SU(6) symmetric case $g_{10}=1$. 

The partial waves in (\ref{eq:decu1p}) were obtained by expanding the
$s$-channel amplitude, aside from spinor factors, in powers of $s$ and
$t$ and then projecting them onto the appropriate Legendre
polynomials.  The same procedure has been applied to $u$-channel
amplitudes, related to $s$-channel ones by crossing, yielding partial
waves expressed in terms of Legendre functions $Q_\ell$.  The
resulting expressions, however, are considerably lengthier than their
octet baryon analogs (\ref{eq:Su})--(\ref{eq:highu}), so we shall omit
them for brevity.

Tree-level BChPT is not sufficient by itself to describe three-flavor
meson-baryon dynamics, in which the large masses of strange hadrons
induce large meson-baryon couplings.  This problem is especially
severe in the $N\overline{K}$ sector, in which strong coupling effects
such as subthreshold resonances render BChPT inapplicable even at
threshold.  We unitarize the tree-level amplitudes with the method
first introduced in \cite{oll99} in the meson sector and extended to
meson--baryon interactions in \cite{mei00,oll01,jid02}.  A technically
detailed explanation of the method can be found in those references.
We shall limit ourselves here to stating the result of the
unitarization of tree-level amplitudes.

Given a set of coupled reaction channels $|B^{a_i}M^{b_i}\rangle
\rightarrow |B^{a_j}M^{b_j}\rangle$, we denote $(f_{\ell\pm})_{ij} =
\pw{\ell\pm}{a_i b_j}{a_j b_j}$ the corresponding tree-level partial
wave matrix.  A solution to the unitarity equation for $T$, resumming
the right-hand cut in the $s$-plane, is given by the partial waves
$(\mathcal{F}_{\ell\pm})_{ij}$ related to $(f_{\ell\pm})_{ij}$ by the
matrix equation,
\begin{equation}
  \label{eq:unitarization}
  \mathcal{F}_{\ell\pm} = \left( I + f_{\ell\pm} \cdot G \right)^{-1}
  \cdot f_{\ell\pm}~,
\end{equation}
where $I$ is an identity matrix, and $G$ is the diagonal matrix
$G_{ij} = g^{a_i b_i} \delta_{ij}$ (no summation over $i$, $j$).  The
``unitarity bubbles'' $g^{ab}$ are given by,
\begin{equation}\label{eq:bubble}
  \begin{split}
  g^{ab}(s) &= \frac{i\mu^\epsilon}{(2\pi)^d} \int d^dk \frac{1}
  {(k^2-\mb^2+i 0)((k+p_T)^2 - m_a^2+i 0)}\\
  &= \frac{1}{16\pi^2} \left\{ a^{ab} +
      \log\left(\frac{\mb_a^2}{\mu^2}\right) +
     \frac{s-\mb_a^2+\mm_b^2}{2s} \log\left( \frac{\mm_b^2}{\mb_a^2}
     \right) + \frac{w(s,\mb_a^2,\mm_b^2)}{2s} \right. \times \\
  &\times \left\{\left[  \log\left( \frac{\mb_a^2 - \mm_b^2 -s -
          w(s,\mb_a^2,\mm_b^2)}{2s} -i 0  \right) -
      \log\left( \frac{\mb_a^2 - \mm_b^2 -s +
          w(s,\mb_a^2,\mm_b^2)}{2s} + i 0  \right)
        \right]
    \rule{0pt}{16pt}\right\}~,
  \end{split}
\end{equation}
with $w(x,y,z)$ defined in appendix \ref{sec:kin}.  The loop function
$g^{ab}$ was computed in (\ref{eq:bubble}) in dimensional
regularization.  The subtraction constants $a^{ab}$, depending on the
renormalization scale $\mu$, are taken as free parameters in each
isospin channel.  Variations in $\mu$ can be offset by a redefinition
of $a^{ab}$ \cite{mei00}. 

For numerical computations we use physical meson and baryon masses,
and coupling constants $D=0.80\pm0.01$ and $F=0.46\pm 0.01$
\cite{clo93} (see also \cite{bor99,rat99}).  Following
\cite{oll01,jid02}, below we set $\mu=630$ MeV.  With our notation for
initial and final momenta, we define the ``lab'' frame by 
$\vec{p}$=0, the ``lab$'$'' frame by $\vec{p}\,'$=0, and the CM
frame by $\vec{p}+\vec{q}\,$=0.

\section{Results for $\boldsymbol{p K^-}$ scattering}
\label{sec:pKm}

In this section we discuss our results for the reactions $p K^-
\rightarrow B^{a'} M^{b'}$.  Among the ten possible final states $p
K^-$, $n\overline{K}^0$, $\Lambda\pi^0$, $\Sigma^0\pi^0$,
$\Sigma^-\pi^+$, $\Sigma^+\pi^-$, $\Lambda\eta$, $\Sigma^0\eta$,
$\Xi^0 K^0$, $\Xi^- K^+$, only the first six are open for initial
meson momentum $0\leq q_\mathrm{lab} \lesssim 300$ MeV.  Following
\cite{oll01} (see also \cite{ben02,jid02,ose98}), we unitarize the
tree-level partial waves with the method of \cite{mei00,oll01,jid02}
taking into account all ten intermediate meson-baryon states.  This is
justified by the fact that in lowest-order BChPT those states are
degenerate.  Notice, however, that physical masses must be used in the
unitarity relations, which are exact.

In the computation of observables such as cross sections and
polarizations we include the $S$, $P_{1/2}$ and $P_{3/2}$ partial
waves arising from the contact meson-baryon interaction and $s$- and
$u$-channel octet baryon exchange, as well as $s$-channel decuplet
baryon exchange, in fully relativistic BChPT as discussed above.  At
the low energies considered here the contribution of $D$ and higher
partial waves is negligible.  We do not expect $u$-channel decuplet
baryon exchange to be significant either, as we have checked to be the
case for several processes in this sector.

In the isospin limit the unitarity bubble $g^{ab}$ contains six
subtraction constants $a^{ab}$ (with $ab=$ $N\overline{K}$,
$\Sigma\pi$, $\Lambda\pi$, $\Sigma\eta$, $\Lambda\eta$, $\Xi K$), that
must be fitted to data.  Also extracted from data are the values of
the common pseudoscalar octet decay constant $f$ and the bare masses
$M_\Sigma$ and $M_\Lambda$ entering the $s$-channel contribution to
the $P_{1/2}$ wave, and $M_{\Sigma^*}$ entering the $s$-channel
contribution to the $P_{3/2}$ wave.  The inclusion of baryon-decuplet
interactions introduces two additional free parameters, the off-shell
parameter $Z$ (or equivalently, $\kappa = 1/2 + Z$) and the coupling
$g_{10}$.  Whereas the latter may be made flavor-dependent, we prefer
not to do so to avoid an excessive proliferation of parameters, and
also because numerical checks suggest that no significant improvement
of our fits is obtained by doing so.  Physical masses are used in all
numerical computations.

In all cases we fit data up to incident momenta
$q_\mathrm{lab}\sim$300---320 MeV, above which $D$-wave effects start
being important.  Since fit parameters are highly correlated, we
cannot vary them arbitrarily.  Instead, in order to check the
stability of, and to estimate the uncertainty in, our results we
fitted the experimental data in several different ways, as described
below.

The data included in our fits consists of the threshold
branching fractions $\gamma$, $R_c$ and $R_n$ \cite{tov71,nov78},
customarily included in $p K^-$ coupled channel analyses \cite{sie88,
  lut00, oll01, ose98, jid02}, total cross sections for the six open
reactions channels, differential cross sections for $p K^-\rightarrow
N\overline{K}$, and the first two Legendre moments of the CM frame
differential cross sections and spin asymmetries for $pK^-\rightarrow
\Sigma\pi, \Lambda\pi$.  We do not include in our fits data on kaonic
hydrogen energy levels and their widths \cite{kek,bee05}.  The most
recent of those experiments \cite{bee05} has obtained precise energy
and width measurements whose phenomenological consequences are
discussed in \cite{mei04z,bor05x,oll05x,bor06z,oll06y,bor05y,oll06z}. 

We present results from four different fits to data.  First, we
consider a non-relativistic HBChPT approximation in which only the
contact interaction contributes to the $S$ wave, only contact
interactions and $s$-channel octet baryon exchange contribute to the
$P_{1/2}$ wave, and only the pole term of $s$-channel decuplet baryon
exchange is included in the $P_{3/2}$ wave.  There is no dependence on
$\kappa$ in this approximation, $g_{10}$ is set to 1 and the values of
the other parameters are taken from \cite{jid02}.  We denote this
approximation as ``\xjor'' below.  In all other fits we consider the
partial waves in the fully relativistic theory as described above.

Second, we consider a fit to the fractions $\gamma$, $R_c$ and $R_n$,
total cross-section data for the six channels open at $q_\mathrm{lab}
\lesssim 300$ MeV, and differential cross-section data for $p K^-
\rightarrow p K^-, n \overline{K}^0$.  We obtain the parameters,
\begin{equation}
  \label{eq:xIV}
  f = 106.90,            ~
  a^{N\overline{K}} = -1.98, ~
  a^{\Lambda\pi} = -3.55,    ~
  a^{\Sigma\pi} = -1.65,     ~
  a^{\Lambda\eta} = -2.30,   ~
  a^{\Sigma\eta} = -5.53,    ~
  a^{\Xi K} =  0.20,
\end{equation}
with bare masses $M_\Sigma = 1237$, $M_\Lambda = 1258$,
$M_{\Sigma^*} = 1320$ MeV, and decuplet parameters $\kappa = 0.4$
and $g_{10}   = 1.10$.  We refer to this fit as ``\xIV''
below. 

Third, adding to the data in fit \xIV\ the Legendre moments data we
obtain the parameter set
\begin{equation}
  \label{eq:xII}
                  f = 101.44,    ~
  a^{N\overline{K}} =  -1.94,    ~
     a^{\Lambda\pi} =  -3.69,    ~
      a^{\Sigma\pi} =  -1.47,    ~
    a^{\Lambda\eta} =  -2.11,    ~
     a^{\Sigma\eta} =  -4.04,    ~
          a^{\Xi K} =   5.02,
\end{equation}
with bare masses $M_\Sigma = 1297$, $M_\Lambda = 1258$, $M_{\Sigma^*}
= 1301$ MeV, and decuplet parameters $\kappa = 0.45$ and $g_{10} =
1.26$.  We refer to this fit as ``\xII'' below.

Lastly, we repeated the series of fits leading to \xII\ with a modified
$\chi^2$ function.  In order to prevent $\chi^2$ from being dominated
by the more numerous total cross section data we weighted each data
set by the inverse number of points.  With this, we obtained the
slightly modified parameter set,
\begin{equation}
  \label{eq:xI}
                  f = 102.42  ,    ~
  a^{N\overline{K}} =  -1.95  ,    ~
     a^{\Lambda\pi} =  -3.88  ,    ~
      a^{\Sigma\pi} =  -1.49  ,    ~
    a^{\Lambda\eta} =  -2.13  ,    ~
     a^{\Sigma\eta} =  -4.05  ,    ~
          a^{\Xi K} =   5.13  ,
\end{equation}
with bare masses $M_\Sigma = 1127$, $M_\Lambda = 1258$, $M_{\Sigma^*}
= 1299$ MeV, and decuplet parameters $\kappa = 0.33$ and $g_{10} =
1.28$.  We refer to this fit as ``\xI'' below. 

Some remarks are in order regarding these parameter sets.  First of
all, we stress here that we are fitting data from many different
experiments, which are not necessarily fully compatible among them.
The values of $f$ in all fits are intermediate between $f_\pi$ and
$f_K$, as expected \cite{jid02}. The constant $g_{10}$ measures the
coupling of $\Sigma$(1385), the only decuplet state exchanged in the
$s$ channel.  In all cases we get $g_{10} \gtrsim 1$, by up to
$\sim$25\%; these results should be taken with caution since, as per
the caveat about data above, our fits are not meant as a determination
of decuplet couplings.  The range of values spanned by the subtraction
constants $a^{ab}$ in fit \xIV\ can be narrowed if higher $\chi^2$
values are accepted, but we prefer to reproduce the data as accurately
as possible.  Inclusion of Legendre-moment data in fits \xI\ and \xII\
leads to a noticeable dispersion in the values of $a^{ab}$.  We have
found that attempting to narrow their range unacceptably worsens the
resulting fits.

Very tight fits to branching fractions and total cross sections are
obtained if only those data are considered, but the resulting fits
lead to a poor description of differential cross section data,
especially Legendre moments for $pK^-\rightarrow \Sigma\pi,
\Lambda\pi$.  Inclusion of all differential cross section data, as in
\xI, \xII, leads to slightly looser but still acceptable fits to total
cross sections.  For fit \xI\ we get the branching fractions $\gamma$
= 2.221 (2.36$\pm$ 0.04), $R_c$ = 0.650 (0.664$\pm$0.011), $R_n$ =
0.208 (0.189$\pm$ 0.015), in good agreement with the experimental
values \cite{tov71,nov78} quoted in parentheses.  Fits \xII, \xIV\
lead to very similar values, while \xjor\ yields a somewhat better
agreement with data \cite{jid02}.  Also, as a verification we computed
the $\Sigma\pi$ mass distribution from the squared isoscalar amplitude
for $N\overline{K} \rightarrow \Sigma\pi$.  We obtain a stable result
for the resonance $\Lambda(1405)$, independently of the fit we use the
resonance peak appears at 1419 MeV with a width of $28$ MeV,
consistent with the results for this channel from the very detailed
study \cite{jid03}.

\begin{figure}[p]
  \centering
  \scalebox{0.775}{\includegraphics{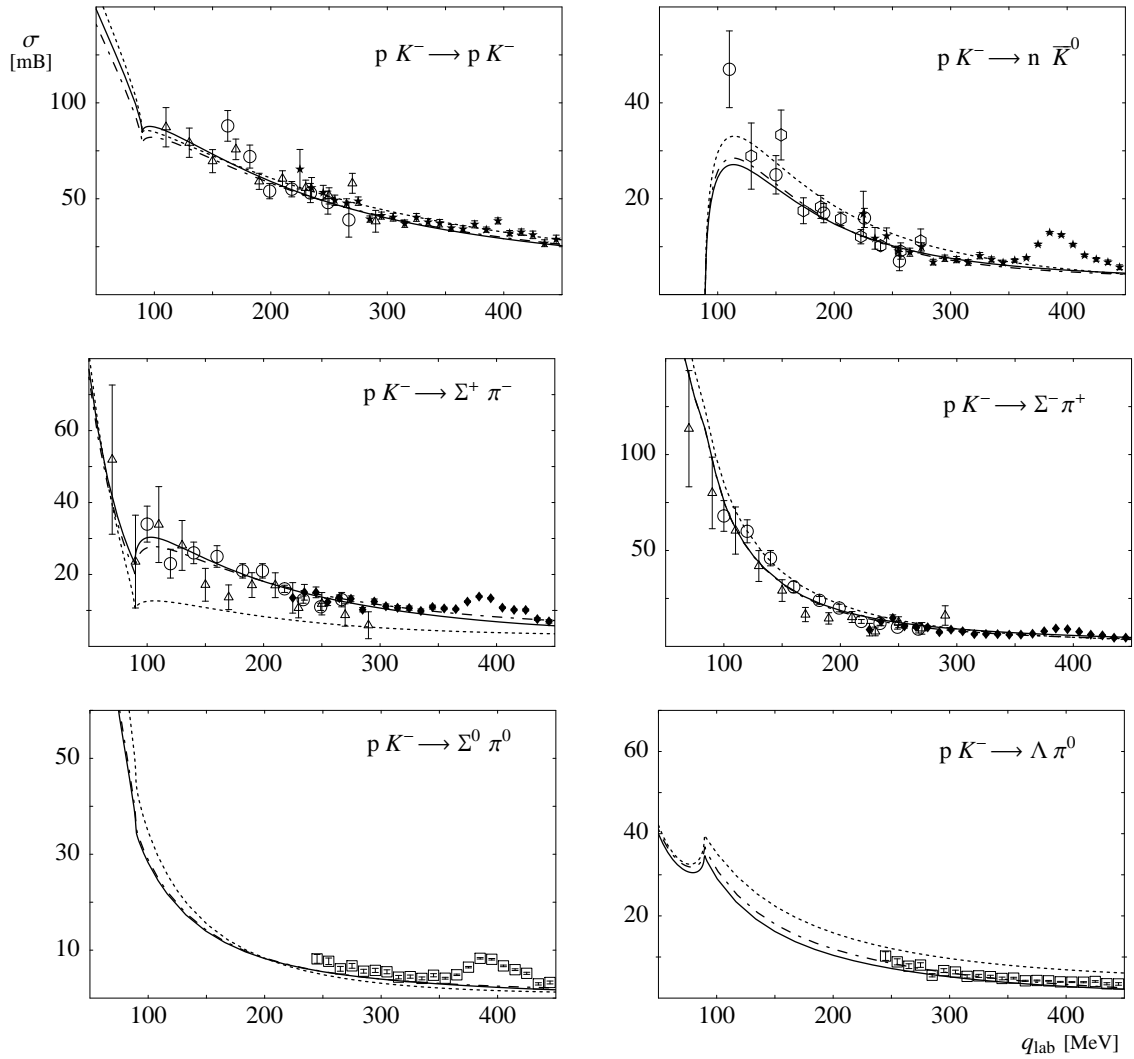}}
\caption{Total cross sections for $pK^-$ scattering.  Solid lines:
  results from fit \xI, dot-dashed lines: fit \xIV, dotted lines: fit
  \xjor.  See text for explanation of fits.  Data from
  \cite{mas75,sak65,eva83,cib82,mas76,ban81}.}
  \label{fig:fig1}
\end{figure}

Our results for total cross sections are shown in fig.\
\ref{fig:fig1}.  As expected, fit \xIV\ gives slightly better results
than \xI.  The contributions to the $S$ wave from $s$- and $u$-channel
baryon octet exchange in those fits lead to a significan improvement
over the results from \xjor.  Fig.\ 2 shows our results for
differential cross sections for $p K^- \rightarrow p K^-, n
\overline{K}^0$.  Although the difference between \xIV\ and \xI\ is
more marked in this case, with the latter giving a better description
of data, all three curves shown in the figure provide a good
description of data within experimental errors. The results for
differential cross-section Legendre moments are significantly
different for the three fits shown in fig.\ \ref{fig:fig3}.  As
expected, parameter set \xI\ yields substantially more accurate
results than \xIV, \xjor, which were not fitted to this data.  Results
from fit \xII\ were omitted from these figures for clarity, since they
are virtually identical to those from \xI.

Fig.\ \ref{fig:fig4} shows our results for the Legendre moments
$B_{1,2}$ of the CM-frame spin asymmetry.  Fits \xI\ and \xII\ give a
much better description of data than \xIV\ and \xjor, as expected
since the latter two do not include these data.  The agreement of the
former two with data is excellent for the charged modes
$pK^-\rightarrow \Sigma^\pm \pi^\mp$.  For the moment $B_1/A_0$ in
$pK^- \rightarrow\Lambda\pi^0$ all data points have a positive central
value, whereas all four fits yield slightly negative values.  This
phenomenon is seen also in fig.\ 8 of \cite{lut00}, where results very
similar to our fit \xI\ are obtained with the same experimental data.
Clearly, new data would help resolve this issue.  \footnote{Also, in
  \cite{lut00} the data for $B_1/A_0$ for $pK^- \rightarrow
  \Sigma^0\pi^0$ appear reflected through the horizontal axis.  Such
  reflection would bring our fits \xI\ and \xII\ to a much better
  agreement with data.  Still, given the large errors in these data,
  they have a small influence on fits.}

We show our results for final baryon polarization $\P'_\mathrm{lab'}$
in fig.\ \ref{fig:fig5}.  Only results for fits \xI\ and \xII\ are
shown, since only those give a good description of the spin-asymmetry
data in fig.\ \ref{fig:fig4}.  As seen in the figure, there are small
quantitative differences in the results for $\P'_\mathrm{lab'}$ from
\xI\ and \xII\ for $pK^-\rightarrow p K^-$, $\Sigma^0\pi^0$, and
slightly larger ones in $pK^-\rightarrow \Sigma^+\pi^-$,
$\Lambda\pi^0$.  We take those differences as a rough estimate of the
uncertainty in our results.

In the process $pK^- \rightarrow n\overline{K}^0$ there is a global
sign change in $\P'_\mathrm{lab'}$: at low values of $q_\mathrm{lab}$
$\P'_\mathrm{lab'}$ takes small but negative values over the entire
range of $\cos \theta_{CM}$, which change to positive values around
$q_\mathrm{lab}\sim 100$ MeV.  We do not plot polarization values
below $\sim$1\%, so that sign change is not visible in the figure.  As
shown there, the difference between curves \xI\ and \xII\ is
larger in this case than in the processes mentioned above. 

A more marked sign change of $\P'_\mathrm{lab'}$ is seen in the plots
for $pK^- \rightarrow \Sigma^-\pi^+$, with small positive
polarizations at small $q_\mathrm{lab}$ turning negative at
$q_\mathrm{lab}\sim$ 150--200 MeV.  As is clear from the figure, this
type of rapid evolution of $\P'_\mathrm{lab'}$ with energy is
sensitive to the parameters and not predicted reliably within the
present approximation.  Taking into account both higher order
corrections in the perturbative amplitudes and higher partial waves in
the unitarized ones would allow us to take into account higher energy
data (up to $q_\mathrm{lab}\sim$ 500 MeV if $D$ waves are included
\cite{lut00}).  This would result in tighter constraints on the energy
dependence of amplitudes and, presumably, reduce the uncertainties in
the extrapolation to lower energies.

\section{Results for $\boldsymbol{p K^+}$ scattering}
\label{sec:pKp}

No $s$-channel interactions or resonance formation are present in $p
K^+$ scattering. Associated production of $\Delta$ occurs about $350$
MeV above threshold in the CM frame, with the rest mass of $\Delta K$
at the $\Delta$ peak at $q_\mathrm{lab} \sim 800$ MeV, so it can be
safely neglected in the range $q_\mathrm{lab} \lesssim 300$ MeV
considered here.  We do take into account $u$-channel $\Sigma^0(1385)$
exchange, though its contributions can be expected to be small.  We
use the $S$, $P_{1/2}$ and $P_{3/2}$ partial waves from contact and
$u$-channel meson--baryon interactions, and from $u$-channel
decuplet baryon exchange, at tree level as input for one-channel
unitarization.
 
The unitarized partial waves have four free parameters, the common
pseudoscalar meson decay constant $f$, the subtraction constant
$a^{NK}$, and the couplings $\kappa$ and $g_{10}$ from decuplet baryon
interactions, which are obtained from fits to differential and total
cross section data \cite{cam74} for $q_\mathrm{lab} < 330$ MeV.  Our
best fit to these data is obtained for $f = 106.04$ MeV, $a^{NK} =
-1.60$, $\kappa = 0.4$, and $g_{10} = 0.79$.  This value of $f$ is
slightly larger than that obtained in the previous section.  Almost
equally good fits are obtained for values of $f$ about 10 MeV higher
or lower, with corresponding changes in the other parameters.  

Results for differential cross sections for 145 $\leq q_\mathrm{lab}
\leq$ 325 MeV are shown in fig.\ \ref{fig:fig6}.  We corrected our
UBChPT partial waves for final-state Coulomb interaction with the
Coulomb amplitudes of \cite{gia71} (see also \cite{lut00}).  The
agreement with data is excellent.  Also shown in the figure is the
differential cross section from Coulomb-uncorrected UBChPT amplitudes,
and from tree-level level amplitudes including neither decuplet baryon
nor Coulomb interactions.  As expected, the contribution from
$u$-channel baryon decuplet interactions is very small at these
energies: setting $g_{10}$=0 leads to barely visible changes in the
figure.  Total cross-section data is shown in fig.\ \ref{fig:fig7}.

Results for polarization are shown in fig.\ \ref{fig:fig8}.  As
expected, the effect of the Coulomb interaction is to erase
polarization, and it is stronger at lower $q_\mathrm{lab}$ and in the
forward direction.  Also shown in the figure are the UBChPT results
without baryon-decuplet contributions, and the result from
leading-order BChPT (without both decuplet baryon and Coulomb
interactions) for $\P'_\mathrm{lab'}$ \cite{bou}.  As seen in the
figure the contribution to $\P'_\mathrm{lab'}$ from $\Sigma^0(1385)$
exchange is small, about $\sim 10$\% of the polarization
value at the higher end of the energy range in the figure and smaller
at lower energies.  The agreement between Coulomb-uncorrected UBChPT
without decuplet baryons and leading-order BChPT results is remarkably
good up to $q_\mathrm{lab} \sim 300$ MeV, where higher-order
correction effects become apparent.  The fact that the UBChPT
polarization is well approximated by the leading-order one at low
energies shows that the former scales with $f$ approximately as
$1/f^2$, as the latter does exactly.

\section{Results for $\boldsymbol{N\pi}$ scattering}
\label{sec:npi}

$N\pi$ scattering at low energies, $q_\mathrm{lab} \lesssim$ 100 MeV,
is well described by one-loop HBChPT \cite{fet01}. Inclusion of
$\Delta$ degrees of freedom extends the applicability of HBChPT up to
$q_\mathrm{lab} \lesssim$ 300 MeV, as shown in \cite{fet00}.
Unitarization, and inclusion of the $N(1440)$ and $\rho$ meson, allows
further extension to $q_\mathrm{lab} \lesssim$ 400 MeV \cite{mei00}.
In the context of fully relativistic BChPT the $N\pi\Delta$
interaction has been studied at one-loop level in \cite{hac05}.  No
detailed phenomenological analyses of $N\pi$ processes such as
\cite{fet01,fet00} have been given in one-loop BChPT yet.  In this
section we apply the unitarized tree-level BChPT partial waves given
above to $N\pi$ scattering.  At near-threshold energies, tree-level
UBChPT is necessarily less precise than a full one-loop treatment.  As
discussed below, however, it yields a very good description of $N\pi$
data up to $q_\mathrm{lab} \lesssim$ 350 MeV, with many fewer free
parameters.

For simplicity we restrict ourselves here to $p\pi^+$ scattering.
Besides the elastic process, also $p\pi^+ \rightarrow \Sigma^+ K^+$
should in principle be considered.  Unitarity corrections arising from
the couplings of those channels, separated by more than $600$ MeV in
the CM frame, are expected to be negligible at the energies considered
here.  Indeed, if that were not so the applicability of BChPT to
$N\pi$ scattering would be called into question.  We therefore apply a
one-channel unitarization procedure to the elastic $p\pi^+$ amplitude,
analogous to the treatment of $pK^+$ scattering above.

We unitarize the tree-level partial waves $S$, $P_{1/2}$, $P_{3/2}$
computed from contact, $s$- and $u$-channel meson-baryon diagrams, and
$s$- and $u$-channel $\Delta$ exchange.  The free parameters are
determined from a fit to differential cross-section and analyzing
power data in the region 150 $\leq q_\mathrm{lab} \leq$ 330 MeV,
around and below the $\Delta$ resonance.  A series of fits was
performed in order to check the stability of our results, with only
small differences found among them.  The parameter $\kappa$, in
particular, was found to be around 0.4 in all cases, with small
fluctuations \cite{mei00}.  Using only differential cross section
data, the best fit is found for
\begin{equation}
  \label{eq:fitIppip}
  f = 78.81\; \mathrm{MeV}~,\quad
  a^{N\pi} = -1.54~,\quad
  \kappa = 0.38~,\quad
  g_{10} =  0.82~,
\end{equation}
with a bare mass $M_\Delta =$ 1285 MeV.  We refer to this fit as ``I''
below.  For elastic processes lab-frame analyzing power is equal to
lab$'$-frame polarization, as can be easily checked (see, however,
\cite{ohl72} for an exhaustive discussion of polarization
observables).  Including analyzing power data in our fits we find,
\begin{equation}
  \label{eq:fitIIppip}
  f = 78.39\; \mathrm{MeV}~,\quad
  a^{N\pi} = -1.77~,\quad
  \kappa = 0.37~,\quad
  g_{10} =  0.81~,
\end{equation}
with a bare mass $M_\Delta =$ 1299.5 MeV.  We refer to this fit as
``II''.  The values for $f$ obtained in these fits is lower than the
one from fits to $pK^\pm$ data above, which may be explained by the
fact that only $N\pi$ interactions are present in this case.  With
respect to the value of $g_{10}$ the same caveats as in sect.\
\ref{sec:pKm} apply here.  The $\Delta$ resonance peak in the
$P_{3/2}$ wave is located at $\sim$1226 MeV, with a Breit-Wigner width
of $\sim$110 MeV, in both cases I and II.

The resulting differential cross sections are plotted in fig.\
\ref{fig:fig9} for $q_\mathrm{lab} \lesssim$ 350 MeV.  Only a
representative sample of data is shown in the figure, including only
the energies with the largest number of data points, since the entire
set comprises twenty different values of $q_\mathrm{lab}$. The
agreement with data is excellent, given the small experimental errors
especially in the most recent experiments \cite{pav01}.  For these
data the difference between I and II is barely visible in the figure.
Polarization results are shown in fig.\ \ref{fig:figb} to be in very
good agreement with data.\footnote{Notice that we define polarization
  in lab frame with respect to $\vec{q}\wedge\vec{p}\,'$ $\propto$
  $-\vec{q}\wedge\vec{q}\,'$ \cite{bou}, so a sign-flip has been
  applied to the experimental data in fig.\ \ref{fig:figb}.}  The
difference between the two fits is noticeable here at peak
polarizations.  Aside from that small difference, the result I is
remarkable since polarization data were not included in that fit.

If we neglect decuplet baryons, the results from UBChPT agree quite
precisely with those of leading order BChPT \cite{bou} up to
$q_\mathrm{lab} \sim 300$ MeV, similarly to the case of $p K^+$
scattering discussed above.  Inclusion of $\Delta$ degrees of freedom
leads to large effects.  A comparison of results with and without
$\Delta$ was given in \cite{bou}, where only the pole term in the
$P_{3/2}$ wave was taken into account.  The polarizations shown in
fig.\ \ref{fig:figa} are substantially larger than those in
\cite{bou}.  The latter are reproduced in the present approach if,
firstly, we increase the value of $f$ (here extracted from data) to
the larger nominal value used in \cite{bou}.  And, second, if we
restrict the $\Delta$-exchange contribution to $s$-channel $P_{3/2}$
wave as done in that ref.  The contribution of $s$-channel $\Delta$
exchange to the $S$ partial wave, in particular, is responsible for
the enhancement of polarization with respect to \cite{bou} not
accounted for by the change in $f$.  As shown in fig.\ \ref{fig:figa},
the contribution of $u$-channel $\Delta$ exchange to
$\P'_\mathrm{lab'}$ is comparatively small, except near the backward
direction at the highest energies plotted there.  Together with the
results of \cite{bou}, those presented in this figure give a complete
picture of $\Delta$-resonance contributions to nucleon polarization at
low energies.


\section{Final remarks}
\label{sec:finrem}

In this paper we computed the polarization of the final-state baryon
in its rest frame in low-energy meson--baryon scattering with
unpolarized initial state, in the unitarization framework of
\cite{mei00,oll01,jid02} applied to fully relativistic leading-order
BChPT.

For $pK^-$ scattering very good agreement is found between UBChPT and
unpolarized cross section data, confirming (and to some extent,
improving) several previous analyses \cite{oll01,jid02,ben02}.  The
agreement with CM-frame spin-asymmetry data is also good, though
somewhat less certain, as seen in fig.\ \ref{fig:fig4}.  We attribute
this to the scarcity of such data and the rather large errors in some
of the existing measurements.  On the other hand, it is precisely spin
observables that put the tightest constraints on fit parameters, since
the clearest discrimination among our different fits is provided by
fig.\ \ref{fig:fig4}.

The results in fig.\ \ref{fig:fig5} show large values of polarization
for $pK^-\rightarrow \Sigma\pi$, $\Lambda\pi$ even at relatively low
energies ($q_\mathrm{lab} \sim 150$ MeV).  It is interesting also that
sizable polarizations are obtained for the elastic and quasielastic
processes $pK^-\rightarrow pK^-$, $n\overline{K}^0$ at the upper end
of the energy range considered in fig.\ \ref{fig:fig5}.  The stability
of the results, as indicated by the two curves in fig.\
\ref{fig:fig5}, varies widely for the different processes.  Whereas
the results for $pK^-$, $\Sigma^0\pi^0$, $\Sigma^+\pi^-$,
$\Lambda\pi^0$ final states seem reliable, those for $n\overline{K}^0$
are less stable, and for $\Sigma^-\pi^+$ largely inconclusive.  New,
more precise data on unpolarized differential cross sections at low
energies would certainly be most helpful to reduce uncertainties.
Alternatively, as mentioned in sect.\ \ref{sec:pKm}, extending the
theoretical computation to fit the more abundant higher-energy data
(along the lines of \cite{lut00}) could lead to a more tightly
constrained extrapolation to lower energies. 

For $pK^+$ scattering excellent agreement between our UBChPT results
and differential cross-section data is shown in fig.\ \ref{fig:fig6}.
Significant ($\gtrsim$ 10\%) values of polarization are found at the
upper end of the energy range of fig.\ \ref{fig:fig8}, even after
correcting for Coulomb interaction.  The agreement of the UBChPT
result for polarization with that of leading-order BChPT \cite{bou} in
fig.\ \ref{fig:fig8} is quite remarkable for $q_\mathrm{lab} \lesssim
300$ MeV.  Since those calculations were carried out by very different
methods, their agreement provides a cross-check for both.
Experimental data at these energies would obviously be most
desirable. 

In the case of $p\pi^+$ scattering also excellent agreement is found
with differential cross-section data in the $\Delta$ peak region, as
(partially) shown in fig.\ \ref{fig:fig9}.  Such agreement is
remarkable given the high precision of some of the data sets used, and
the fact that our results are based on unitarized tree-level
amplitudes.  More importantly, our results for polarization are in
very good agreement with experimental data, as seen in fig.\
\ref{fig:figa}, which provides a non-trivial validation to our
calculations.  

Interestingly, a comparison of fig.\ \ref{fig:figa} with the results
of \cite{bou} shows that the non-resonant $S$-wave contribution from
$s$-channel $\Delta$ exchange has a strong influence on polarization,
suggesting that precise polarization data could discriminate among
different $\Delta$-$N$-$\pi$ interaction models.  As also shown in the
figure, the contribution to polarization from $u$-channel $\Delta$
exchange, though significant,  is relatively small.

We hope that the results presented here may motivate new measurements
at low energies, or re-analyses of existing data, leading to new and
more precise experimental information on polarization observables
especially in the strange sector.
\section*{Acknowledgement}

I thank the first referee for providing helpful information on
analyzing powers and polarization.

\appendix

\section{Kinematics}
\label{sec:kin}

In this appendix we gather some kinematical definitions used 
throughout the paper.  We introduce the notation
\begin{equation}
  \label{eq:omega}
  \omega(x,y,z) = (x^2+y^2+z^2-2xy-2xz-2yz)^\frac{1}{2}
  = (x-(\sqrt{y}+\sqrt{z})^2)^\frac{1}{2}
  (x-(\sqrt{y}-\sqrt{z})^2)^\frac{1}{2}~. 
\end{equation}
The function $\omega$ appears frequently in relativistic kinematics
(\emph{e.g.,} in the center of mass frame $|\vec{p}| =
\omega(s,\mb_a^2, \mm_{b}^2)/(2\sqrt{s})$). The Mandelstam invariants
for the process $| B^{a}(p,\sigma) M^b(q)\rangle \longrightarrow |
B^{a'}(p',\sigma') M^{b'}(q')\rangle$ are
\begin{equation}
  \label{eq:mandelstam}
  s=(p+q)^2=(p'+q')^2~,
  \quad
  t=(p-p')^2=(q-q')^2~,
  \quad
  u=(p-q')^2=(p'-q)^2~,
\end{equation}
with $s+t+u = \mb_a^2 + \mb_{a'}^2 + \mm_{b}^2 + \mm_{b'}^2~$.
The physical region for the process is defined by the inequalities
\begin{equation}
  \label{eq:physreg}
  s_\sss{\mathrm{th}}\leq s~,
  \quad
  \tmin \leq t \leq \tmax~,  
  \quad
  \umin \leq u \leq \umax~,
\end{equation}
where,
\begin{equation}
  \label{eq:kinlim}
  \begin{aligned}
    s_\sss{\mathrm{th}} &= \max\left\{(\mb_a + \mm_b)^2, (\mb_{a'} +
      \mm_{b'})^2 \right\}\\
    t_{\substack{\sss{\mathrm{max}}\\[-2pt]\sss{\mathrm{min}}}} &=
    -\frac{1}{2s} \left( \rule{0pt}{10pt} s^2 - s (\mb_a^2 +
      \mb_{a'}^2 + \mm_b^2 + \mm_{b'}^2 ) + (\mb_a^2 - \mm_b^2)
      (\mb_{a'}^2 - \mm_{b'}^2) \right) \\
    &\quad \pm \frac{1}{2s} \omega(s,\mb_a^2,\mm_b^2)
      \omega(s,\mb_{a'}^2,\mm_{b'}^2),\\
    u_{\substack{\sss{\mathrm{max}}\\[-2pt]\sss{\mathrm{min}}}} &=
    -\frac{1}{2s} \left( \rule{0pt}{10pt} s^2 - s (\mb_a^2 +
      \mb_{a'}^2 + \mm_b^2 + \mm_{b'}^2 ) - (\mb_a^2 - \mm_b^2)
      (\mb_{a'}^2 - \mm_{b'}^2) \right) \\
    &\quad \pm \frac{1}{2s}
    \omega(s,\mb_a^2,\mm_b^2) \omega(s,\mb_{a'}^2,\mm_{b'}^2).\\    
  \end{aligned}
\end{equation}

\begin{figure}[p]
  \centering
  \scalebox{0.65}{\includegraphics{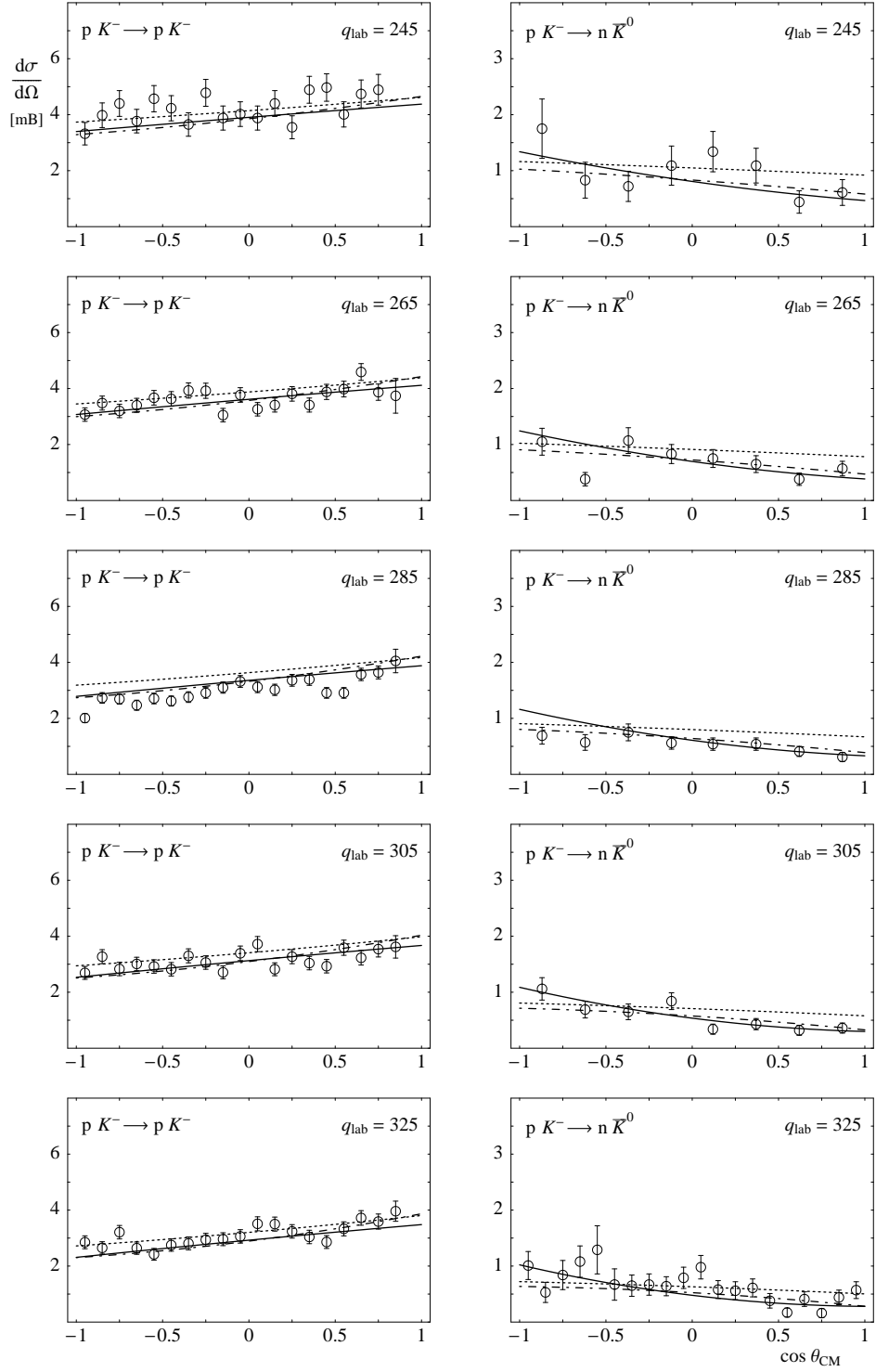}}
  \caption{Differential cross sections for $pK^-$ scattering.  Solid,
    dot-dashed and dotted lines as in fig.\ \ref{fig:fig1}. Data from
    \cite{mas76}.}
  \label{fig:fig2}
\end{figure}
\begin{figure}[p]
  \centering
  \scalebox{0.65}{\includegraphics{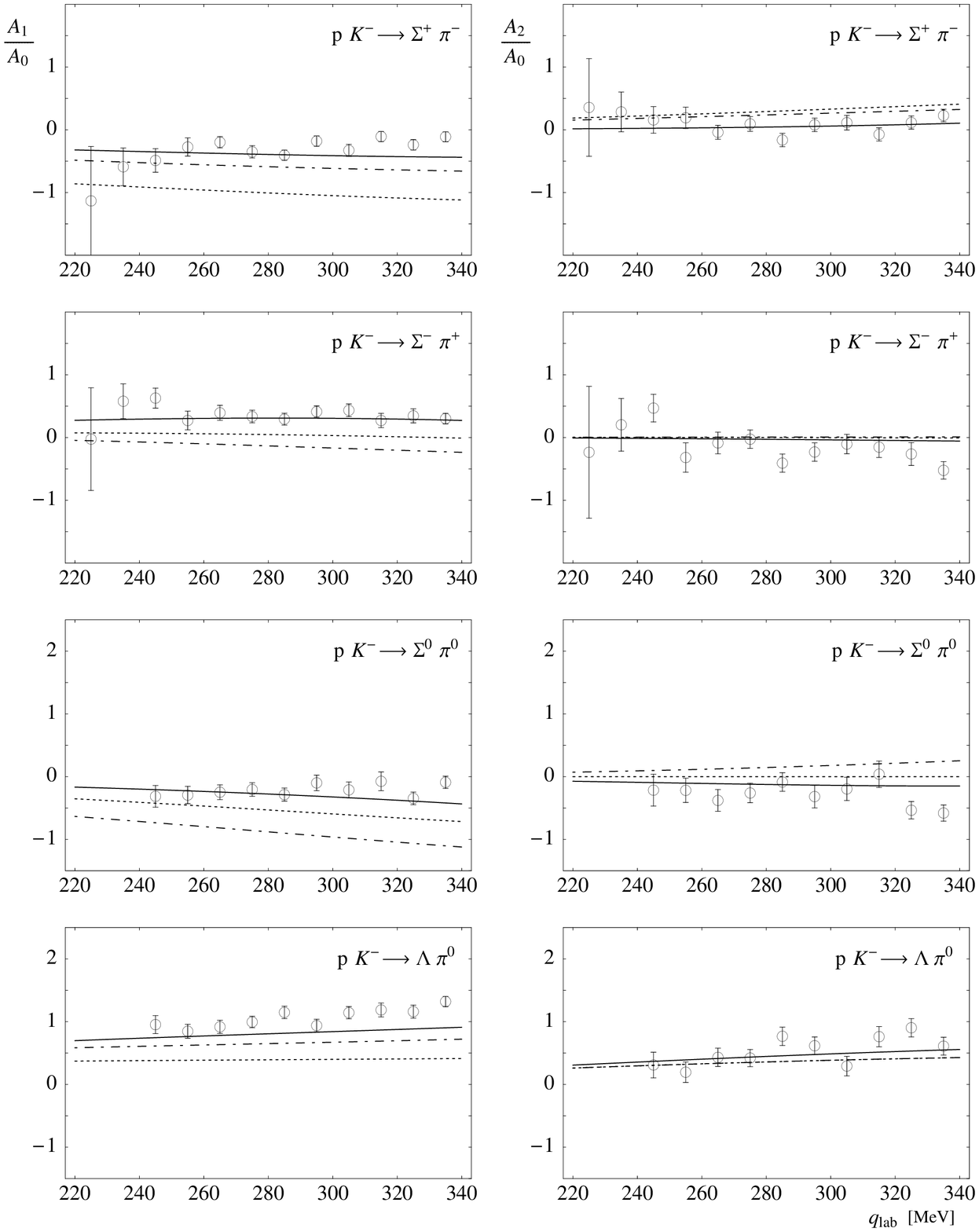}}
  \caption{Legendre moments $A_{1,2}$ for $pK^-$ scattering.  Solid,
    dot-dashed and dotted lines as in fig.\ \ref{fig:fig1}. Data from
    \cite{mas75,ban81}.}
  \label{fig:fig3}
\end{figure}
\begin{figure}[p]
  \centering
  \scalebox{0.65}{\includegraphics{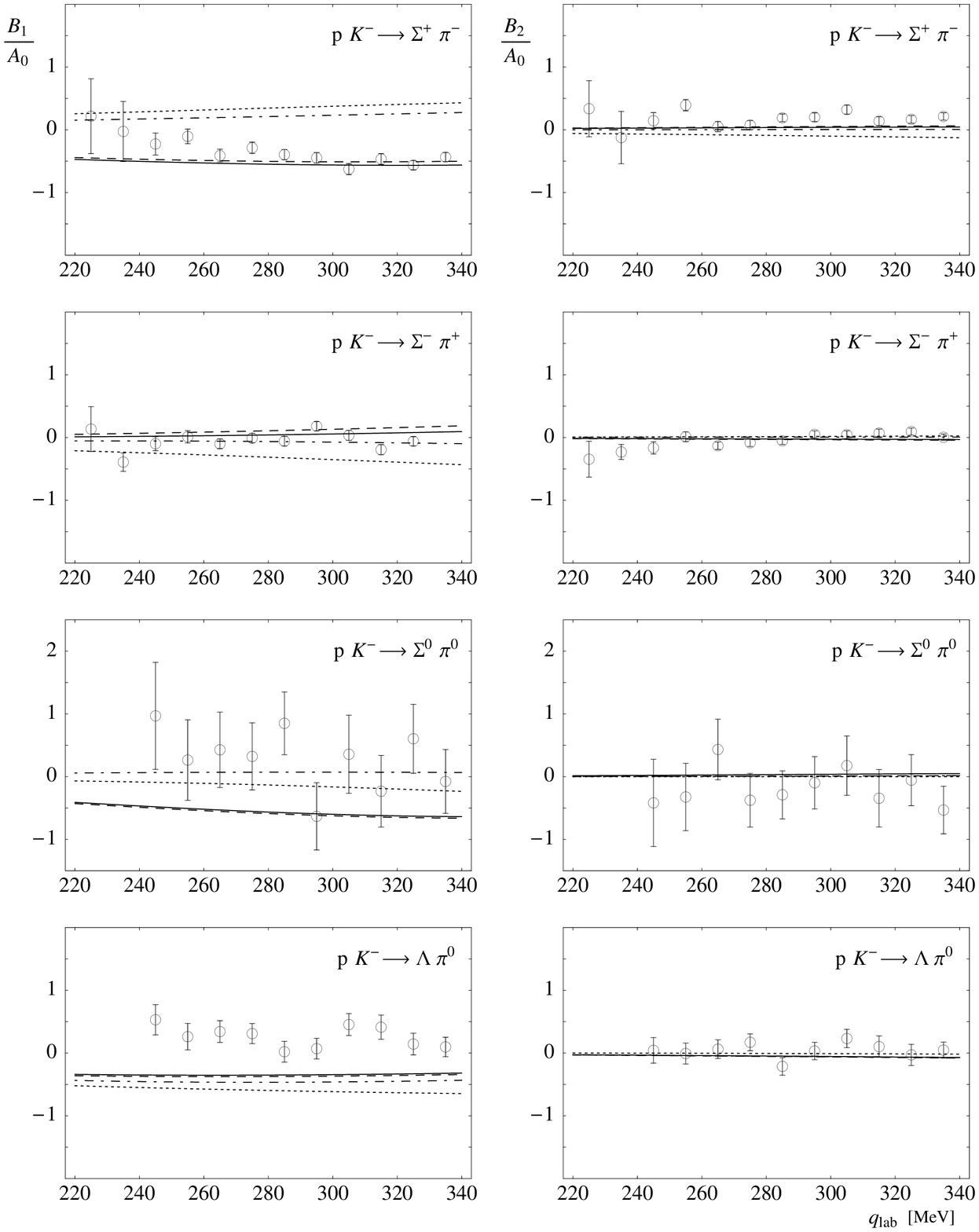}}
\caption{Legendre moments $B_{1,2}$ for $pK^-$ scattering.  Solid,
    dot-dashed and dotted lines as in fig.\ \ref{fig:fig1}. Dashed
    lines: fit \xII. Data from \cite{mas75,ban81}.}
  \label{fig:fig4}
\end{figure}
\begin{figure}[p]
  \centering
  \scalebox{0.75}{\includegraphics{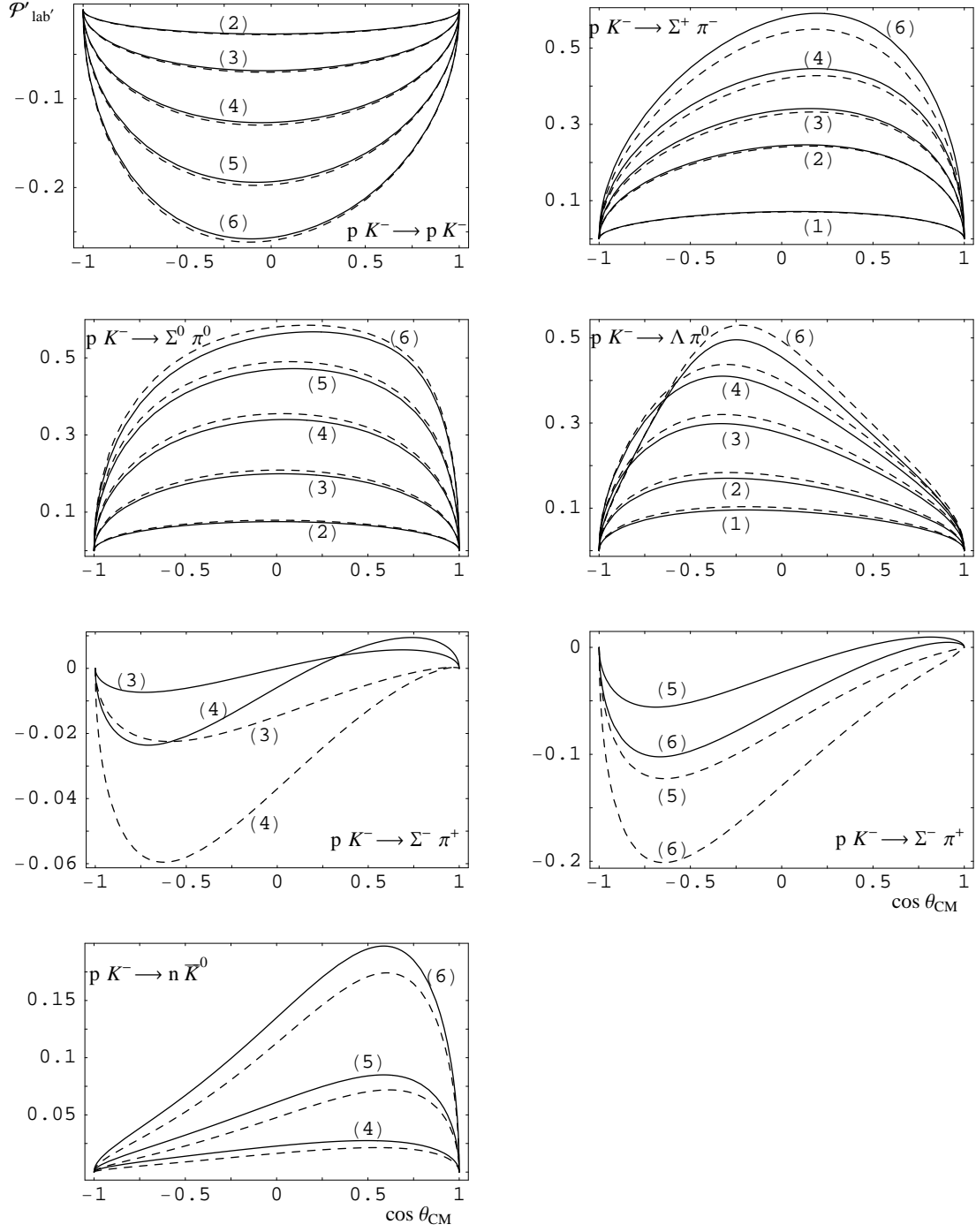}}
  \caption{UBChPT results for the polarization of the final baryon in
    its rest frame in $pK^-$ unpolarized scattering, as a function of
    center-of-mass scattering angle.  Solid lines: parameter set \xI,
    dashed lines: parameter set \xII\ (see text for an explanation of
    parameter sets).  Curves (1)---(6) (not all shown in some plots
    for clarity): $q_\mathrm{lab}=$ 50, 100, 150, 200, 250, 300 MeV,
    resp.}
  \label{fig:fig5}
\end{figure}
\begin{figure}[p]
  \centering
  \scalebox{0.75}{\includegraphics{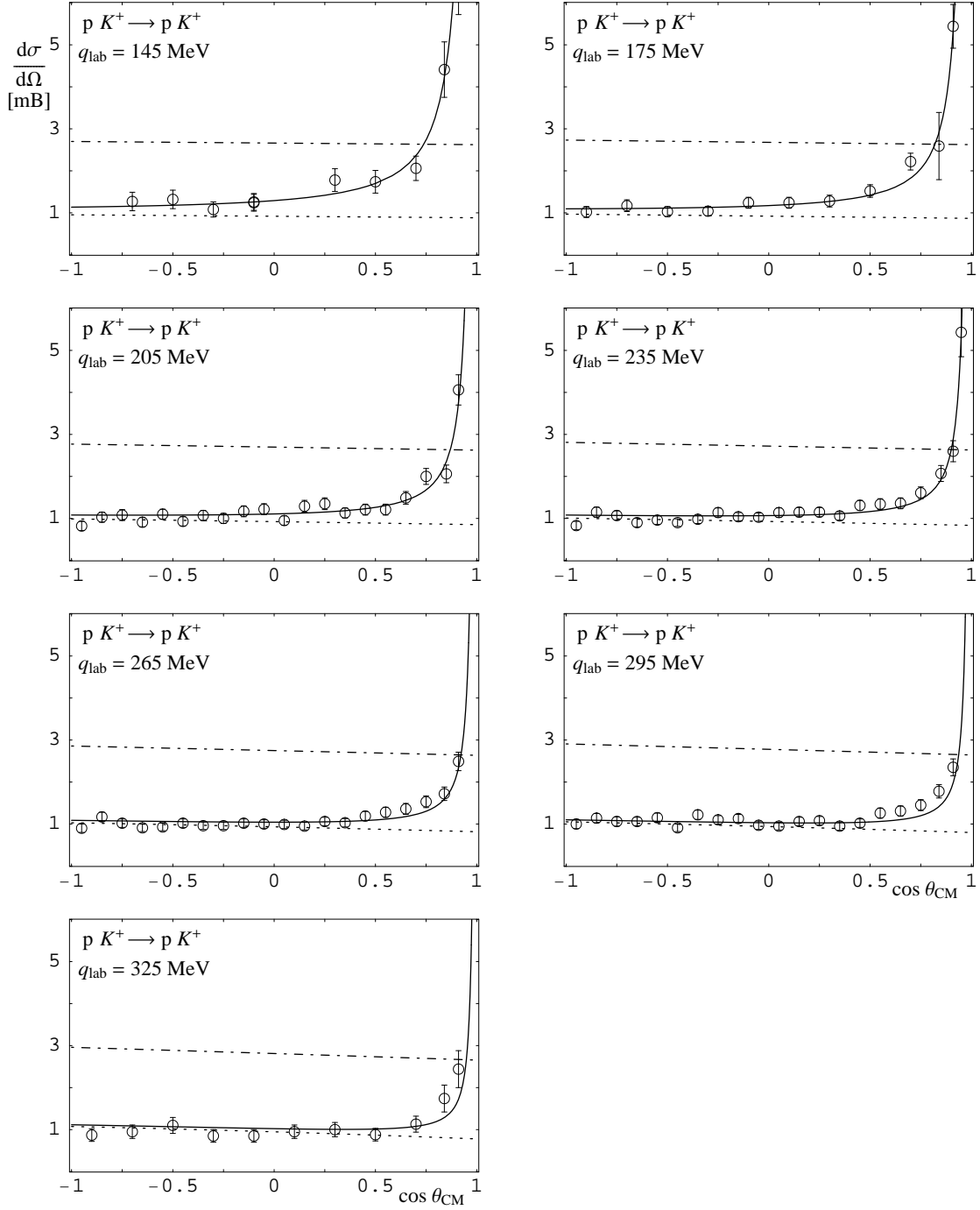}}
  \caption{Differential cross section for $pK^+$ scattering. Solid
    lines: UBChPT results corrected for Coulomb interaction, dotted
    lines: uncorrected UBChPT results, dot-dashed lines: tree-level
    BChPT results. Data from \cite{cam74}.}
  \label{fig:fig6}
\end{figure}
\begin{figure}[p]
  \centering
  \scalebox{0.75}{\includegraphics{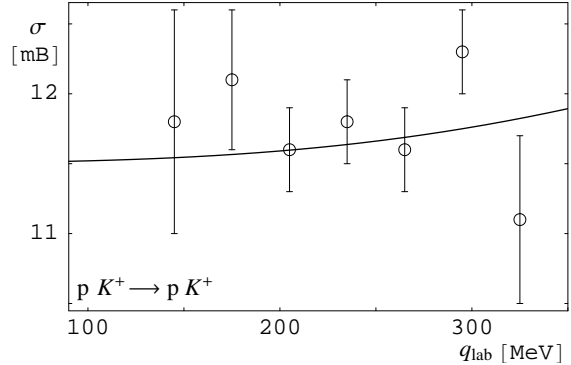}}
  \caption{Total cross section for $pK^+$ scattering.  Solid line:
    UBChPT result.  Data corrected for Coulomb interaction from
    \cite{cam74}.} 
  \label{fig:fig7}
\end{figure}
\begin{figure}[p]
  \centering
  \scalebox{0.75}{\includegraphics{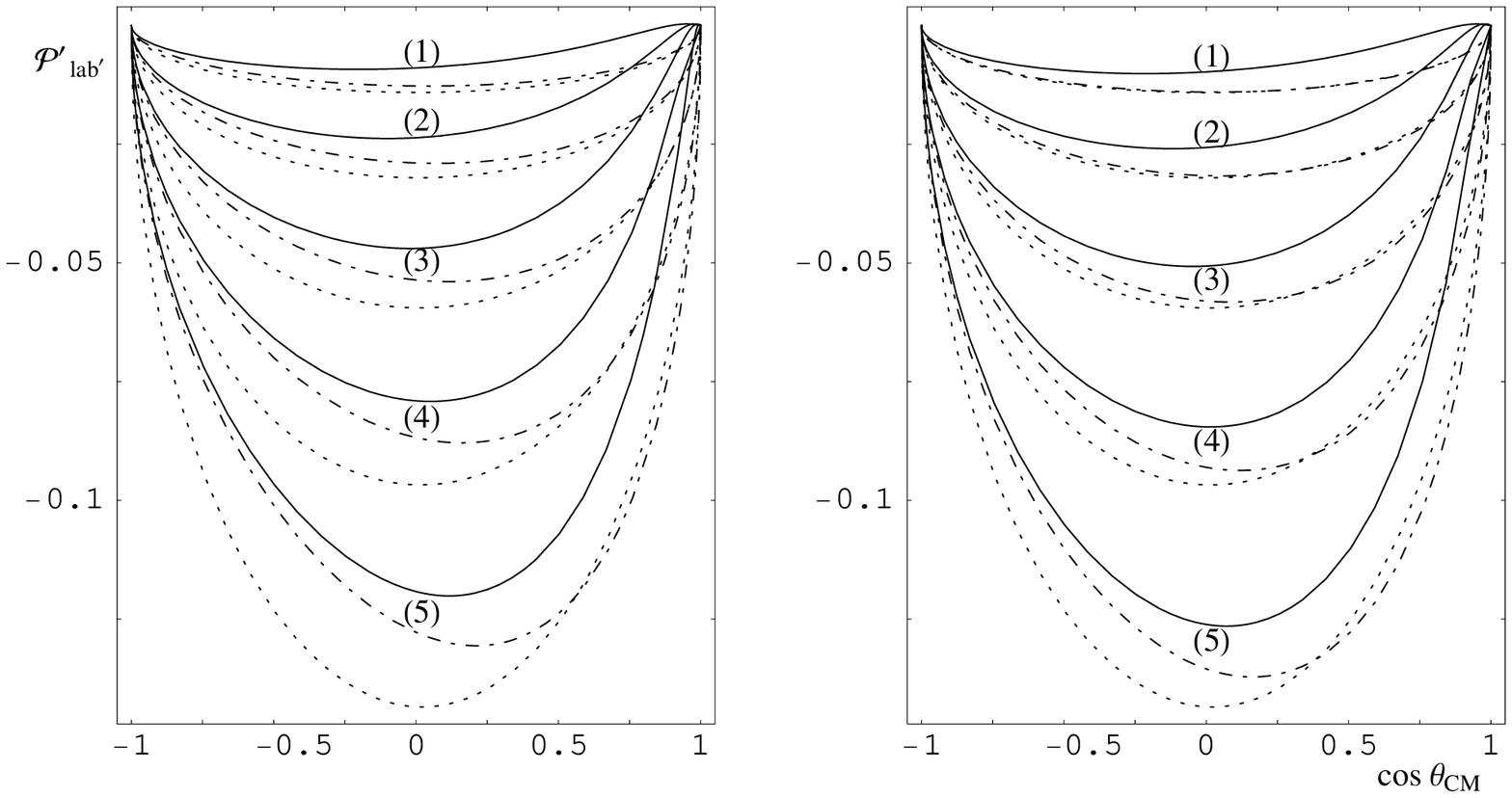}}
  \caption{UBChPT results for the polarization of the final baryon in
    its rest frame in $pK^+$ unpolarized scattering, as a function of
    center-of-mass frame scattering angle.  Left panel: complete
    UBChPT including baryon decuplet interactions. Right panel: UBChPT
    with baryon decuplet interactions switched off.  Solid lines:
    UBChPT with Coulomb interaction, dot-dashed lines: UBChPT without
    Coulomb corrections, dotted lines: leading-order BChPT calculation
    without both baryon decuplet and Coulomb interactions \cite{bou}.
    Curves (1)---(5): $q_\mathrm{lab} =$ 150, 200, 250, 300, 350 MeV,
    resp.}
  \label{fig:fig8}
\end{figure}
\begin{figure}[p]
  \centering
\rotatebox{-90}{\scalebox{0.7}{\includegraphics{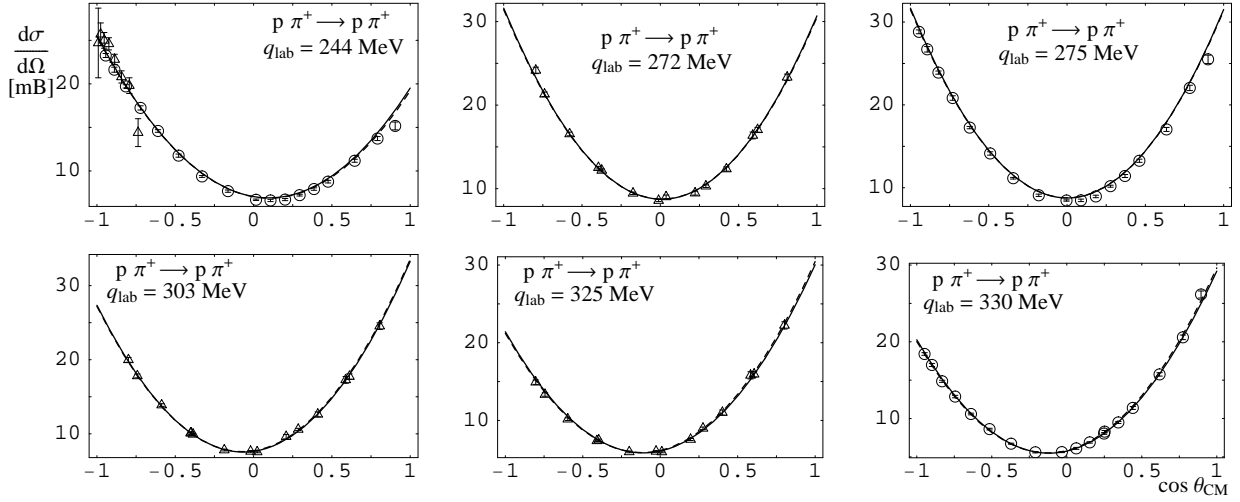}}}
\caption{Differential cross section for $p\pi^+$ scattering. Data from
  \cite{pav01,sta80,bus73,rit83}.}
   \label{fig:fig9}
\end{figure}
\begin{figure}[p]
  \centering
\rotatebox{-90}{\scalebox{0.72}{\includegraphics{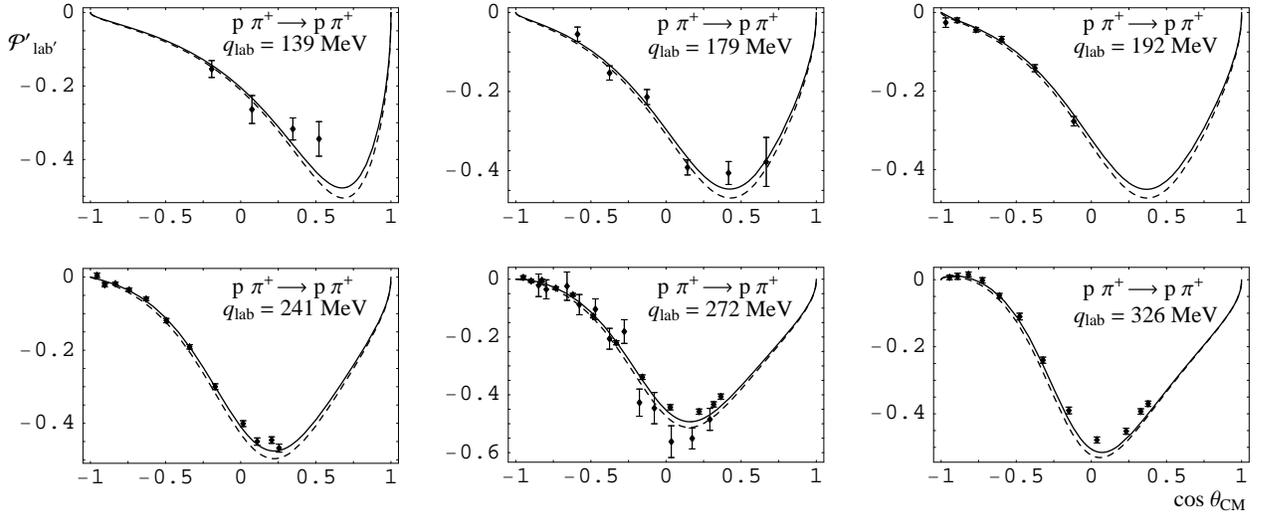}}}
\caption{UBChPT results for the polarization of the final baryon in
  its rest frame in $p\pi^+$ unpolarized scattering, as a function of
  center-of-mass frame scattering angle.  Dashed lines: fit I, solid
  lines: fit II.  Data from \cite{sev89,ams76,mei04}, after sign
  flipping. (See main text for explanation of fits and sign
  convention.)  For data from \cite{mei04}, statistical and systematic
  errors added in quadrature.}
  \label{fig:figb}
\end{figure}
\begin{figure}[p]
  \centering
\scalebox{0.65}{\includegraphics{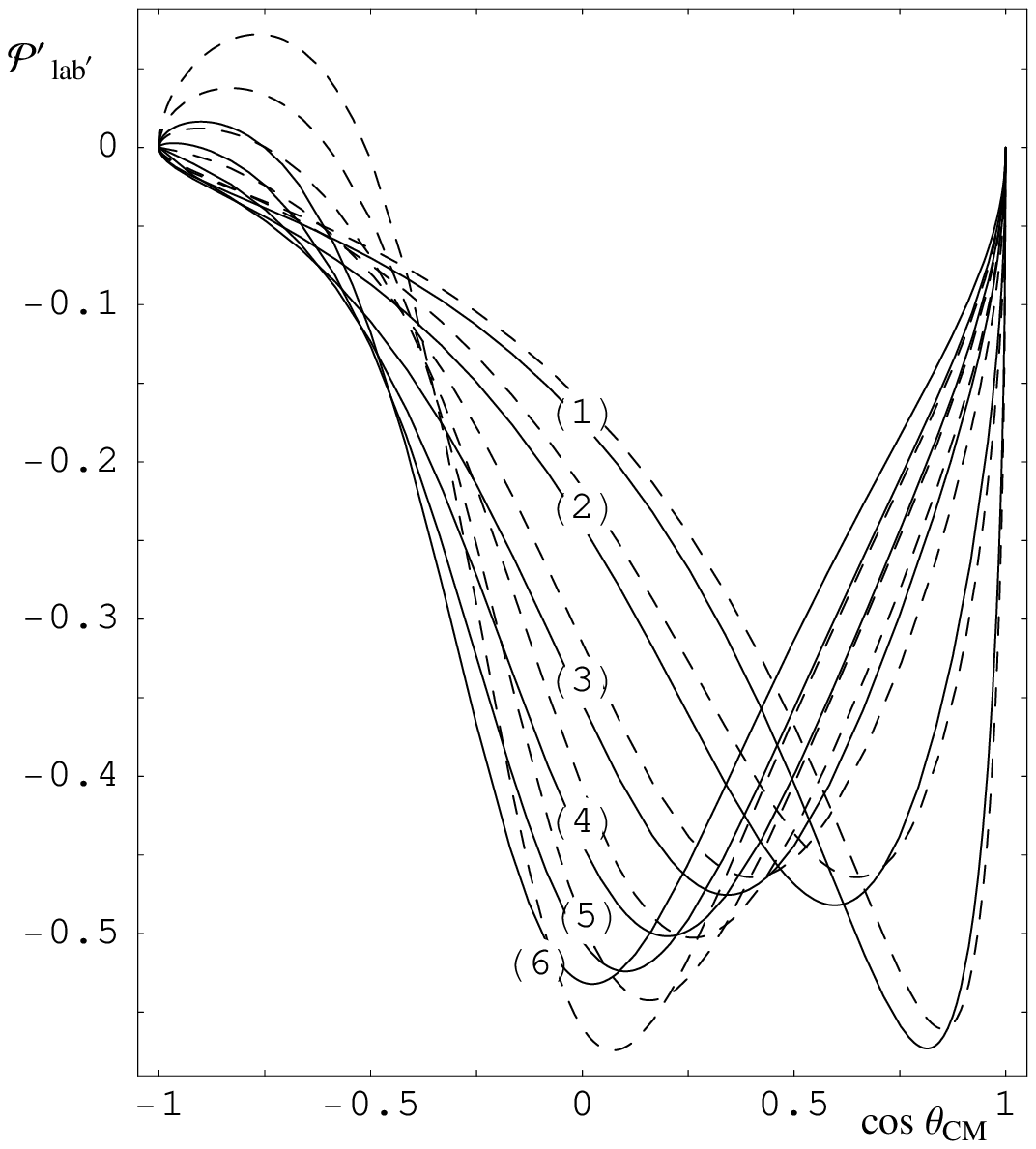}}
\caption{UBChPT results for the polarization of the final baryon in
  its rest frame in $p\pi^+$ unpolarized scattering, as a function of
  center-of-mass scattering angle.  Solid lines: full computation with
  both $s$- and $u$-channel $\Delta$--exchange diagrams, with
  parameters from fit I.  Dashed lines: same computation, without
  $u$-channel $\Delta$-exchange diagram.  Curves (1)---(6):
  $q_\mathrm{lab} =$ 125, 150, 200, 250, 300, 350 MeV, resp.}
  \label{fig:figa}
\end{figure}

\begin{thebibliography}{99}
%
\bibitem{ber07a}V.\ Bernard, \emph{Chiral Perturbation Theory and
    Baryon Properties,} arXiv:0706.0312. 
%
\bibitem{sch05}S.\ Scherer, M.\ R.\ Schindler, \emph{A Chiral
    Perturbation Theory Primer,} lectures given at the European Centre
  for Theoretical Studies in Nuclear Physics and Related Areas,
  Trento, Italy, 2005; arxiv:hep-ph/0505265.
%
\bibitem{sie88} P.\ Siegel, W.\ Weise, Phys.\ Rev.\ C \textbf{38}
  (1988) 2221.
%
\bibitem{kai95} N.\ Kaiser, P.\ B.\ Siegel, W.\ Weise, Nucl.\ Phys.\ A
  \textbf{594} (1995) 325.
%
\bibitem{kai97} N.\ Kaiser, T.\ Waas, W.\ Weise, Nucl.\ Phys.\ A
  \textbf{612} (1997) 297.
%
\bibitem{ose98}E.\ Oset, A.\ Ramos, Nucl.\ Phys.\ A \textbf{635}
  (1998) 99.
%
\bibitem{lut00} M.\ Lutz, E.\ Kolomeitsev, Nucl.\ Phys.\ A
  \textbf{700} (2002) 193. 
%
\bibitem{bor02} B.\ Borasoy, E.\ Marco, S.\ Wetzel, Phys.\ Rev.\ C
  \textbf{66} (2002) 055208.
%
\bibitem{oll99} J.\ A.\ Oller, E.\ Oset, Phys.\ Rev.\ D \textbf{60}
  (1999) 074023; J.\ A.\ Oller, E.\ Oset, J.\ R.\ Pel\'aez, Phys.\
  Rev.\ D \textbf{59} (1999) 074001.
%
\bibitem{mei00} U.\ G.\ Meissner, J.\ A.\ Oller, Nucl.\ Phys.\ A
  \textbf{673} (2000) 311.
%
\bibitem{oll01}J.\ A.\ Oller, U.\ G.\ Meissner, Phys.\ Lett.\ B
  \textbf{500} (2001) 263.
%
\bibitem{jid02}D.\ Jido, E.\ Oset, A.\ Ramos, Phys.\ Rev.\ C
  \textbf{66} (2002) 055203.
%
\bibitem{jid03} D.\ Jido, J.\ A.\ Oller, E.\ Oset, A.\ Ramos, U.-G.\
  Meissner, Nucl.\ Phys.\ A \textbf{725} (2003) 181.
%
\bibitem{oll06z} J.\ A.\ Oller, Eur.\ Phys.\ J.\ A \textbf{28} (2006)
  63.
%
\bibitem{bou} A.\ Bouzas, Int.\ J.\ Mod.\ Phys.\ E, to appear.
%
\bibitem{ben02}E.\ Oset, A.\ Ramos, C.\ Bennhold, Phys.\ Lett.\ B
  \textbf{527} (2002) 99, Erratum:  Phys.\ Lett.\ B \textbf{530}
  (2002) 260. 
%
\bibitem{don94}J.\ F.\ Donoghue, E.\ Golowich, B.\ R.\ Holstein,
  \emph{Dynamics of the Standard Model,} Cambridge Univ.\ Press, New 
  York, 1994. 
%
\bibitem{gas84}J.\ Gasser, H.\ Leutwyler, Ann.\ Phys.\ (N.Y.)
  \textbf{158} (1984) 142.
%
\bibitem{gas85a}J.\ Gasser, H.\ Leutwyler, Nucl.\ Phys.\ B
  \textbf{250} (1985) 465.
%
\bibitem{kra90}A.\ Krause, Helv.\ Phys.\ Acta \textbf{63} (1990) 3.
%
\bibitem{fri04}M.\ Frink, U.\ G.\ Meissner, J.\ High Energy Phys.\
  \textbf{0407} (2004) 028. 
%
\bibitem{oll06a}J.\ A.\ Oller, J.\ Prades, M.\ Verbeni, J.\ High Energy
  Phys.\  \textbf{0609} (2006) 079,\\ Erratum: {arxiv:hep-ph/0701096}.
%
\bibitem{gas88}J.\ Gasser, M.\ E.\ Sainio, A.\ Svarc, Nucl.\ Phys.\
  \textbf{B 307} (1988) 779.
%
\bibitem{abram} M.\ Abramowitz, I.\ Stegun, ``Handbook of Mathematical
  Functions,'' Dover Pub., New York, 1972. 
%
\bibitem{ber95} V.\  Bernard, N.\  Kaiser, U.-G. Meissner, Int.\ J.\
  Mod.\ Phys.\ E \textbf{4} (1995) 193.
%
\bibitem{ben89} M.\ Benmerrouche, R.\ M.\ Davidson, N.\ C.\
  Mukhopadhyay, Phys.\ Rev.\ C \textbf{39} (1989) 2339.
%
\bibitem{ols78} M.\ G.\ Olsson, E.\ T.\ Osypowski, E.\ H.\ Monsay,
  Phys.\ Rev.\ D \textbf{17} (1978) 2938.
%
\bibitem{pas07} V.\ Pascalutsa, M.\ Vanderhaeghen, S.\ N.\ Yang,
  Phys.\ Rept.\ 437 (2007) 125.
%
\bibitem{leb94} R.\ F.\ Lebed, Nucl.\ Phys.\ B \textbf{430} 1994 295.
%
\bibitem{clo93}F.\ E.\ Close, R.\ G.\ Roberts, Phys.\ Lett.\ B
  \textbf{316} (1993) 165.
%
\bibitem{bor99}B.\ Borasoy, Phys.\ Rev.\ D \textbf{59} (1999) 054021. 
%
\bibitem{rat99}P.\ G.\ Ratcliffe, Phys.\ Rev.\ D \textbf{59} (1999)
  014038.
%
%
\bibitem{tov71} D.\ Tovee et al., Nucl.\ Phys.\ B \textbf{33} (1971)
  493. 
%
\bibitem{nov78} R.\ Novak et al., Nucl.\ Phys.\ B \textbf{139} (1978)
  61. 
\bibitem{kek} M.\ Iwasaki et al., Phys.\ Rev.\ Lett.\ \textbf{78}
  (1997) 3067; T.\ M.\ Ito et al., Phys.\ Rev.\ C \textbf{58} (1998)
  2366. 
%
\bibitem{bee05} G.\ Beer et al., Phys.\ Rev.\ Lett.\ \textbf{94}
  (2005) 212302.
%
\bibitem{mei04z} U.-G.\ Meissner, U.\ Raha, A.\ Rusetsky, Eur.\ Phys.\
  J.\ C \textbf{35} (2004) 349.
%
\bibitem{bor05x} B.\ Borasoy, R.\ Nissler, W.\ Weise, Phys.\ Rev.\
  Lett.\ \textbf{94} (2005) 213401.
%
\bibitem{oll05x} J.\ A.\ Oller, J.\ Prades, M.\ Verbeni, Phys.\ Rev.\
  Lett.\ \textbf{95} (2005) 172502.
%
\bibitem{bor06z} B.\ Borasoy, R.\ Nissler, W.\ Weise, Phys.\ Rev.\
  Lett.\ \textbf{96} (2006) 199201.
%
\bibitem{oll06y} J.\ A.\ Oller, J.\ Prades, M.\ Verbeni, Phys.\ Rev.\
  Lett.\ \textbf{96} (2006) 199202.
%
\bibitem{bor05y} B.\ Borasoy, R.\ Nissler, W.\ Weise, Eur.\ Phys.\ J.\
  A\textbf{25} (2005) 79.
%
\bibitem{cam74} W.\ Cameron et al., Nucl.\ Phys.\ B \textbf{78} (1974)
  93.
%
\bibitem{gia71} G.\ Giacomelli et al., Nucl.\ Phys.\ B \textbf{71}
  (1974) 138. 
%
\bibitem{fet01} N.\ Fettes, U.-G.\ Meissner, Nucl.\ Phys.\ A
  \textbf{693} (2001) 693.
%
\bibitem{fet00} N.\ Fettes, U.-G.\ Meissner, Nucl.\ Phys.\ A
  \textbf{679} (2001) 629. 
%
\bibitem{hac05} C.\ Hacker, N.\ Wies, J.\ Gegelia, S.\ Scherer, Phys.\
  Rev.\ C \textbf{72} (2005) 055203. 
%
\bibitem{ohl72} G.\ G.\ Ohlsen, Rep.\ Prog.\ Phys.\ \textbf{35} (1972)
  717. 
%
\bibitem{pav01} M.\ M.\ Pavan et al., Phys.\ Rev.\ C \textbf{64}
  (2001) 064611.
%
\bibitem{mas75} T.\ S.\ Mast et al., Phys.\ Rev.\ D \textbf{11} (1975)
  3078. 
%
\bibitem{sak65} M.\ Sakitt et al., Phys.\ Rev.\ \textbf{139} (1965)
  B719. 
%
\bibitem{eva83} D.\ Evans et al., J.\ Phys.\ G \textbf{9} (1983) 885. 
%
\bibitem{cib82} J.\ Ciborowski et al., J.\ Phys.\ G \textbf{8} (1982)
  13. 
%
\bibitem{mas76} T.\ S.\ Mast et al., Phys.\ Rev.\ D \textrm{14} (1976)
  13.   
%
\bibitem{ban81} R.\ O.\ Bangerter et al., Phys.\ Rev.\ D \textbf{23}
  (1981) 1484. 
%
\bibitem{sta80} A.\ Stanovnik et al., Phys.\ Lett.\ B \textbf{94}
  (1980) 323.
%
\bibitem{bus73} P.\ J.\ Bussey et al., Nucl.\ Phys.\ B \textbf{58}
  (1973) 363. 
%
\bibitem{rit83} B.\ G.\  Ritchie et al., Phys.\ Lett.\ B \textbf{125}
  (1983) 128.
%
\bibitem{sev89} M.\ E.\ Sevior et al., Phys.\ Rev.\ C \textbf{40}
  (1989) 2780.                      
%
\bibitem{ams76} C.\ Amsler et al., Lett.\ Nuov.\ Cim.\ \textbf{15}
  (1976) 209.
%
\bibitem{mei04} R.\ Meier et al., Phys.\ Lett.\ B \textbf{588} (2004)
  155. 
\end{thebibliography}
\end{document}